\documentclass[twocolumn,twocolappendix]{aastex631}

\def\ms{\mbox{$M_{*}$}}
\def\msun{\mbox{\rm M$_{\odot}$}}

\def\agemw{\mbox{$Age_{\rm mw}$}}
\def\agelw{\mbox{$Age_{\rm lw}$}}
\def\zmw{\mbox{$Z_{\rm \star,mw}$}}
\def\zlw{\mbox{$Z_{\rm \star,lw}$}}

\definecolor{darkgreen}{rgb}{0,0.6,0}

\shorttitle{SDSS-IV MaNGA: The MaNDala Sample}
\shortauthors{M. Cano-D\'{\i}az et al.}
\usepackage{graphicx}
\usepackage[caption=false]{subfig}
\usepackage{wrapfig}
\usepackage{longtable}
\usepackage{hyperref}
\usepackage{amsmath}
\usepackage[normalem]{ulem}
\usepackage{comment}

\graphicspath{{./}{figures/}}

\begin{document}

\title{SDSS-IV MaNGA: The MaNGA Dwarf Galaxy Sample Presentation}

\author[0000-0001-9553-8230]{M. Cano-D\'{\i}az}
\affiliation{CONACYT Research Fellow - Instituto de Astronom\'{\i}a, Universidad Nacional Aut\'onoma de M\'exico, Apartado Postal 70-264, CDMX, 04510 Mexico \\}

\author[0000-0001-9601-7779]{H.M. Hern\'andez-Toledo}
\author[0000-0002-0170-5358]{A. Rodr\'{\i}guez-Puebla}
\affiliation{Instituto de Astronom\'{\i}a, Universidad Nacional Aut\'onoma de M\'exico, Apartado Postal 70-264, CDMX, 04510 M\'exico}

\author[0000-0002-9790-6313]{H.J. Ibarra-Medel}
\affiliation{Instituto de Astronom\'ia y Ciencias Planetarias, Universidad de Atacama, Copayapu 485, Copiap\'o, Chile}

\author[0000-0002-3461-2342]{V. \'Avila-Reese}
\author[0000-0002-0523-5509]{O. Valenzuela}
\affiliation{Instituto de Astronom\'{\i}a, Universidad Nacional Aut\'onoma de M\'exico, Apartado Postal 70-264, CDMX, 04510 M\'exico}

\author[0000-0002-7805-6982]{A.E. Medellin-Hurtado}
\affiliation{Licenciatura en Ciencias de la Tierra, Facultad de Ciencias, Universidad Nacional Aut\'onoma de M\'exico, Circuito Exterior S/N, 04510, CDMX, M\'exico}

\author[0000-0001-8694-1204]{J. A. V\'azquez-Mata}
\affiliation{Departamento de F\'isica, Facultad de Ciencias, Universidad Nacional Aut\'onoma de M\'exico, Ciudad Universitaria, CDMX, 04510, M\'exico}
\affiliation{Instituto de Astronom\'ia sede Ensenada, Universidad Nacional Aut\'onoma de M\'exico, Km 107, Carret. Tij.-Ens., Ensenada, 22060, BC, M\'exico\\}

\author[0000-0002-5908-6852]{A. Weijmans}
\affiliation{School of Physics and Astronomy, University of St Andrews, North Haugh, St Andrews KY16 9SS, UK}

\author[0000-0002-3724-1583]{J.J. Gonz\'alez}
\affiliation{Instituto de Astronom\'{\i}a, Universidad Nacional Aut\'onoma de M\'exico, Apartado Postal 70-264, CDMX, 04510 M\'exico}

\author[0000-0003-1083-9208]{E. Aquino-Ortiz}
\affiliation{Instituto de Astrofísica, Pontificia Universidad Católica de Chile, Av. Vicuña Mackenna 4860, 782-0436 Macul, Santiago, Chile.}

\author[0000-0001-7608-5360]{L. A. Mart\'{\i}nez-V\'azquez}
\affiliation{Instituto de Astronom\'{\i}a, Universidad Nacional Aut\'onoma de M\'exico, Apartado Postal 70-264, CDMX, 04510 M\'exico}

\author[0000-0003-1805-0316]{Richard R. Lane}
\affiliation{Centro de Investigación en Astronomía, Universidad Bernardo O'Higgins, Avenida Viel 1497, Santiago, Chile.}







\begin{abstract}
We present the MaNGA Dwarf galaxy, MaNDala, Value-Added-Catalog, VAC, from the final release of the Sloan Digital Sky Survey-IV program. MaNDala consists of 136 randomly selected bright dwarf galaxies with $\ms<10^{9.1}\msun$ and $M_{g}>-18.5$ making it the largest Integral Field Spectroscopy homogeneous sample of dwarf galaxies. We release a photometric analysis of the $g,r$ and $z$ broadband imaging based on the DESI Legacy Imaging Surveys as well as the spectroscopic analysis based on the Pipe3D SDSS-IV VAC. Our release includes the surface brightness (SB), geometric parameters and color profiles, S\'ersic fits as well as stellar population properties (such as, stellar ages, metallicities, star formation histories), and emission lines fluxes within the FOV and the effective radii of the galaxies. We find that the majority of the MaNDala galaxies are star forming late-type galaxies with $\langle{}n_{\text{Sersic,r}}\rangle\sim1.6$ that are centrals (central/satellite dichotomy). MaNDala covers a large range of SB values (we find 11 candidates of ultra diffuse galaxies and 3 compact ones), filling the gap between classical dwarfs and low-mass galaxies in the Kormendy Diagram and in the size-mass/luminosity relation, whichseems to flatten at $10^8<\ms/\msun<10^{9}$ with $\langle{}R_{e,r}\rangle\sim2.7$ kpc. A large fraction of MaNDala galaxies formed from an early low-metallicity burst of SF but also of late SF events from more metal-enriched gas: half of the MaNDala galaxies assembled $50\%$ of their mass at $\langle{}z\rangle>2$, while the last $20\%$ was at $\langle{}z\rangle<0.3$. Finally, a bending of the sSFR-\ms\ relation at $\ms\sim10^{9}\msun$ for the main sequence galaxies seems to be supported by MaNDala.
\end{abstract}




\section{Introduction} \label{sec:intro}

 Mass is a key property of galaxies that plays a fundamental role while understanding their evolution. In particular, it is notable that the distributions exhibited by the population of galaxies in several of their main properties strongly segregate by stellar mass.
 At $\ms\approx 2-3\times 10^{10}$ \msun, galaxies follow a clear bimodal distribution separated into red/passive/early-type and blue/star-forming/late-type populations, but at much larger or smaller masses, strongly dominate the former or the latter respectively.
 The understanding of the physics beyond this mass-dependent segregation is not yet fully achieved \citep[for recent reviews, see e.g.,][]{Somerville-Dave2015,deLucia2019}; specially at the low-mass end, theory and observations appear to be in tension \citep[see e.g.,][]{Avila-Reese-Firmani2011,Leitner2012,Weinmann+2012,Somerville-Dave2015}.
 Low-mass galaxies are challenging to observe due to their low luminosity and surface brightness (SB), and thus, little is known when compared to their massive counterparts. The difficulties to observe them increases notably when their stellar masses gets lower than $\sim 10^9$ \msun, which is the typical threshold to classify dwarf galaxies (DGs) as such.   
 
The slope of the observed galaxy luminosity function at the low-luminosity side is negative; this implies that dwarfs are the most abundant galaxies in the Universe. 

Within the current cosmological paradigm of galaxy formation, the luminosity function, evolution, and internal properties of these galaxies depend on the nature of dark matter, but also on the effects of the baryonic processes, to which the small-scale structures are highly sensitive due to their shallow gravitational potentials. Thus, the study of DGs becomes crucial (i) to probe cosmological models and the nature of dark matter, and (ii) to understand the complex galaxy baryonic processes and evolutionary trends, such as the formation of molecular clouds and stars, including their dependence on metallicity and the UV  background,  and the feedback of stars and supernovae; for some recent reviews, see \citet[][]{Weinberg+2015,Colin+2015,Bullock+2017}, and more references therein.

 Related to item (i), estimations of the inner dynamical mass distribution of dwarfs (the ‘cusp–core’ controversy), and the central stellar densities of massive dwarf satellites (the ‘too-big-to-fail’ controversy) have been used as probes of dark matter type. Regarding item (ii), an important task is to infer the star formation (SF) and metallicity enrichment histories for large dwarf samples, and how they depend on their mass and environment, as well as whether the trend of downsizing in specific SF rate (sSFR) continues or not below $\sim 10^9$ \msun. Both types of studies can greatly benefit from Integral Field Spectroscopy (IFS) observations, which allow us to obtain resolved stellar populations properties and resolved kinematic information from the stellar and ionized gas components of galaxies. 

Most of the previous detailed observational studies on DGs refer to the Local Group  \citep[e.g.,][]{Tolstoy+2009,McConnachie+2012,Weisz+2014} or to nearby clusters \citep[e.g.,][]{Ferrarese+2012,Eigenthaler2018,Venhola+2019}, meaning these studies are constrained to particular environments. To deeply understand and use DGs to study the small-scale challenges mentioned above, we need to explore them much more in detail in different environments, using both (resolved and unresolved) imaging and spectroscopy studies. 

Early efforts of multi-frequency studies of dwarfs beyond the Local Group comprised only some tens of galaxies, for example, \citet[][ACS Nearby Galaxy Survey Treasury Program, ANGST]{Dalcanton+2009}, \citet[][Little THINGS]{Hunter2012},  \citet[][VLA-ANGST]{Ott+2012}, and \citet[][]{McGaugh+2017}.  \citet{Geha2012} provided a valuable spectro-photometric catalog of about $3000$ local field DGs ($z<0.055$) based on the public NASA-Sloan Atlas Catalog \citep[NSA;][]{Blanton2011}. Based on public
multi-wavelength data sets \citet[][]{Karachentsev+2013} compiled a catalog of galaxies in the Local Volume ($<11$ Mpc), which are mostly dwarfs. Based also on public data sets (the SDSS DR7 and other sources), \citet[][]{Ann+2015} confirmed $\sim 2600$ local dwarfs in the Catalog of Visually Classified Galaxies. More recently, efforts have also been made to study dwarfs beyond the Local Group using resolved stellar maps \citep[e.g.,][for dwarfs around the elliptical galaxy NGC 5128]{Crnojevic+2016}. Also, using deep imaging information from multi-wavelength surveys, studies of dwarf satellites around Milky Way analogs \citep[][]{Bennet+2017,Mao+2021,Carlsten+2021}, dwarfs in the field \citep{Tanoglidis+2021}, and dwarf pairs/groups \citep[TiNy Titans, TNT survey;][]{Stierwalt+2015,Stierwalt+2017} were carried out. Based on a recent deep imaging survey aimed to study low surface brightness (SB) features (including DGs) in the outskirts of nearby massive early type galaxies (the Mass Assembly of early Type gaLAxies with their fine Structures, MATLAS), \citet[][]{Habas+2020} presented the sample selection and photometric properties of 2210 candidate dwarfs, while in \citet[][]{Poulain2021}, the structure and morphology of these galaxies were determined.

Using IFS observations and applying the spectral inversion method, based on fits of a composition of single stellar populations (SSPs) models to the spectra, inferences about the global and spatially-resolved archaeological properties of the galaxies and their evolution can be made \citep[for a recent review, see][]{Sanchez2020}. Specifically, global and radial stellar masses and SF histories of galaxies can be derived \citep[e.g.,][]{Perez2013,Ho2016,Ibarra-Medel+16,Sanchez+2019,Neumann2020}, as well as studies about the SF and the processes that quenched them \citep[e.g.,][]{CatalanTorrecilla2017,Schaefer2019,Cano-Diaz+2019,Lacerna+20}. IFS data also allows us to study the nature and effects of the AGNs in galaxies \citep[e.g.,][]{Mingozzi2019,sanchez18a,Wylezalek2018}, as well as their spatially-resolved kinematics \citep[e.g.,][]{Raouf2021,GarmaOehmichen2020,AquinoOrtiz2020}.

In recent years, the first very large IFS galaxy surveys have been completed. The largest is the Mapping Nearby Galaxies at APO \citep[MaNGA;][]{Bundy15}, which has observed $\sim$ 10,000 local galaxies ($z<0.15$) across 1.5 or 2.5 effective radii $R_e$. Although MaNGA has been designed to roughly uniformly cover the stellar mass range of $10^9<\ms/\msun<10^{12}$, a small fraction of galaxies were observed at smaller masses, including those from an ancillary program dedicated to observe dwarfs (P.I. M. Cano-D\'{\i}az). The goal of this paper, which is the first in a series, is to present the sample of MaNGA galaxies with masses $\ms\lesssim 10^9$ \msun, which are mostly bright dwarfs (all retrieved from the final data release, DR17 \citealp{Abdurrouf+2022}). To our knowledge, with 136 galaxies this is the first large sample of DGs with IFS observations 

We present here the selection criteria and basic photometric and spectroscopic characterizations of the sample, named \textbf{MaN}GA \textbf{D}warf G\textbf{ala}xy (MaNDala), using the MaNGA IFS data and multi-band photometric optical images coming from  the DESI Legacy Imaging Surveys \citep{Dey2019}. Using the IFS data, with the Pipe3D code and its recent improvement, pyPipe3D \citep[][S\'anchez, et al. submitted]{Sanchez16a,sanchez18a}, we perform spectral and archaeological analyzes to characterize the level of star-forming activity of the MaNDala galaxies and determine their global mean ages and stellar metallicities (mass- and luminosity-weighted). From the photometric analysis, we obtain one-dimensional radial SB and color profiles, as well as geometric parameters. This analysis contains a wealth of useful information that allows us to review the global structural properties of the MaNDala sample, but also allows us to infer useful diagnostics for the presence of relevant inner structures and for more subtle structures like warps at the outer regions.

The extensive set of results for the MaNDala dwarfs coming from the two complementary data samples mentioned above will be useful for a diversity of studies, in which we intend to explore different aspects of the nature of these galaxies. All of our results will be publicly available in the form of a Sloan Digital Sky Survey IV (SDSS-IV) Value Added Catalog (VAC)\footnote{Please refer to Appendix \ref{AppendixVAC} for further information}.

The structure of the paper is as follows. In section \ref{Data} we describe the photometric and spectroscopic data used for this work. In section \ref{sample}, the sample selection is described. The photometric and spectroscopic analyses are described in section \ref{Analysis}, while their results are reported in section \ref{Results}. Finally, in section \ref{SummaryDiscussion} we give our summary and discussion.

 Throughout this paper we assume a \citet[][]{Chabrier} initial mass function (IMF) and the following cosmology: $H_0 = 70$ km/s/Mpc, $\Omega_{M}$ = 0.3, and $\Omega_{\Lambda}$ = 0.7.

\section {Data}\label{Data}
\subsection{DESI images}\label{Data_desi}

The DESI Legacy Imaging Surveys \citep{Dey2019} are a combination of three imaging surveys that have mapped contiguous areas of the sky in three optical bands ($g$, $r$ and $z$) to depths $\sim$2 mag deeper than the Sloan Digital Sky Survey imaging \citep[SDSS; e.g.][]{Abazajian2009}. The three surveys are (i) the DECam Legacy Survey (DECaLS) using the Blanco 4m telescope and the Dark Energy Camera \citep[DECam;][]{Flaugher2015}, (ii) the Mayall z-band Legacy Survey (MzLS) using the Mosaic3 camera (Dey et al. 2016) at the Mayall Telescope, and (iii) the Beijing-Arizona Sky Survey (BASS) using the Bok 2.3m telescope/90Prime camera at Kitt Peak (Williams et al. 2004).
The primary goal of the Legacy Surveys is to provide targets for the Dark Energy Spectroscopic Instrument \citep[DESI;][]{DESI2016}.  

The present work is based on the ninth release of the Legacy Surveys project (LS DR9) which contains data from all of the individual components of the Legacy Surveys (BASS, DECaLS and MzLS). It was built on DR8 by improving the reduction techniques and procedures used for the Legacy Surveys. The images of the MaNDala DGs were retrieved in the $grz$ bands, specifying pixel scale (0.262 arsec/pix) and size (800$\times$800 pixels), centered on the r.a. and dec positions appropriate for our image post-processing.

We adopt the flux calibration for BASS, MzLS and DECaLS on the AB natural system of each instrument, respectively. Since the brightness of objects are all stored as linear fluxes in units of nanomaggies, we adopted the conversion from linear fluxes to magnitudes as described in the Photometry section of the Data Release Description\footnote{\url{https://www.legacysurvey.org/dr9/description/}}.  Notice that the fluxes can be negative for faint objects and that it was the case for some of our faintest objects. As representative values we take median 5$\sigma$ point source (AB) depths for areas with different numbers of observations in the different regions of DR9 as g = 24.7 mag r = 24.0 mag and z = 23.0 mag\footnote{\url{https://www.legacysurvey.org/}}

\subsection{MaNGA spectroscopic data}\label{Data_manga}

MaNGA \citep{Bundy15} is one of the main projects of the SDSS-IV international collaboration \citep{Blanton2017}. This project used the IFS technique to observe over 10,000 galaxies by the end of its operations in 2020. Data was acquired with a dedicated 2.5 meter telescope at the Apache Point Observatory (APO) \citep{Gunn06}. To observe the main targets this project used Integral Field Units (IFUs) with different fiber bundles, ranging from 19 to 127 fibers, were each fiber has a diameter of 2$\arcsec$ \citep{Drory15}. This observational setup has a spectral coverage ranging from 3600 to 10300 $\AA$ at a resolution of R$\sim$2000 provided by the dual beam BOSS spectrographs \citep{Smee13}. Smaller fiber bundles were used in simultaneous observations along with the main targets, to perform sky subtraction and flux calibrations \citep{Yan16}. A three point dithering strategy was used for all the observations in order to achieve a complete spatial coverage of the sources within the defined apertures \citep[for these and further details about the observing strategy please refer to][]{Law15}.
We used the 3.1.1 version of the MaNGA reduction pipeline \citep{Law16}, which delivers sky subtracted, wavelength and flux calibrated data cubes as final data products.



\section {Sample Selection}\label{sample}

The MaNGA sample \citep{Wake17} consist of galaxies of all morphological types, redshifts in the range: 0.01 $<z<$ 0.15, and stellar masses, \ms, between $10^{9}$ and $10^{12}$ M$_{\odot}$. Even though the MaNGA Survey has limits in \ms, there is a small fraction of galaxies outside them in the final sample. The MaNDala Sample contains galaxies that surpassed the MaNGA lower limit in \ms, but also galaxies that are part of an ancillary program to specifically observe DGs with the MaNGA observational setup.\footnote{\url{ https://www.sdss.org/dr17/manga/manga-target-selection/ancillary-targets/dwarf-galaxies-with-manga/}}

To define our sample we selected all the galaxies within the final MaNGA sample that have \ms$<$ 10$^{9.06}$\msun,\footnote{Galaxies were originally selected with \ms$<$ 10$^{8.75} h^{-2}$\msun.} after which we obtain 152 galaxies. The stellar masses were retrieved from the NASA-Sloan Atlas Catalog (NSA Catalog\footnote{\url{http://www.nsatlas.org}}; \citealp{Blanton2011}), where the \citealp{Chabrier} initial mass function has been used. We used the available masses derived from a S\'ersic fit, and corrected their values to be in units of $h^{-2}$\msun, considering a value of $h$=0.70, 
instead of $h$=1 as reported in that catalog. Then we discarded all galaxies that are brighter than the Large Magellanic Cloud, following a criterion similar to the one described in \citet{BlantonMoustakas2009}. We eliminated all galaxies that have an absolute magnitude in the $g$ SDSS photometric band reported in the NSA Catalog $<$ -18.5, after this cut we end up with 142 galaxies. We finally discarded the objects for which we did not found a complete set of MaNGA data products, imaging data was not optimal or were suspected to be stars. This final cleanse of the sample reduced it to a final sample of 136 galaxies.

These 136 galaxies conform the first version of the SDSS-IV VAC named MaNDala (V1.0), which is part of the seventeenth data release (DR17) of the SDSS collaboration, whose details are reported in the Appendix \ref{AppendixVAC}.

\begin{figure}
\centering

\subfloat{%
  \includegraphics[width=1\columnwidth]{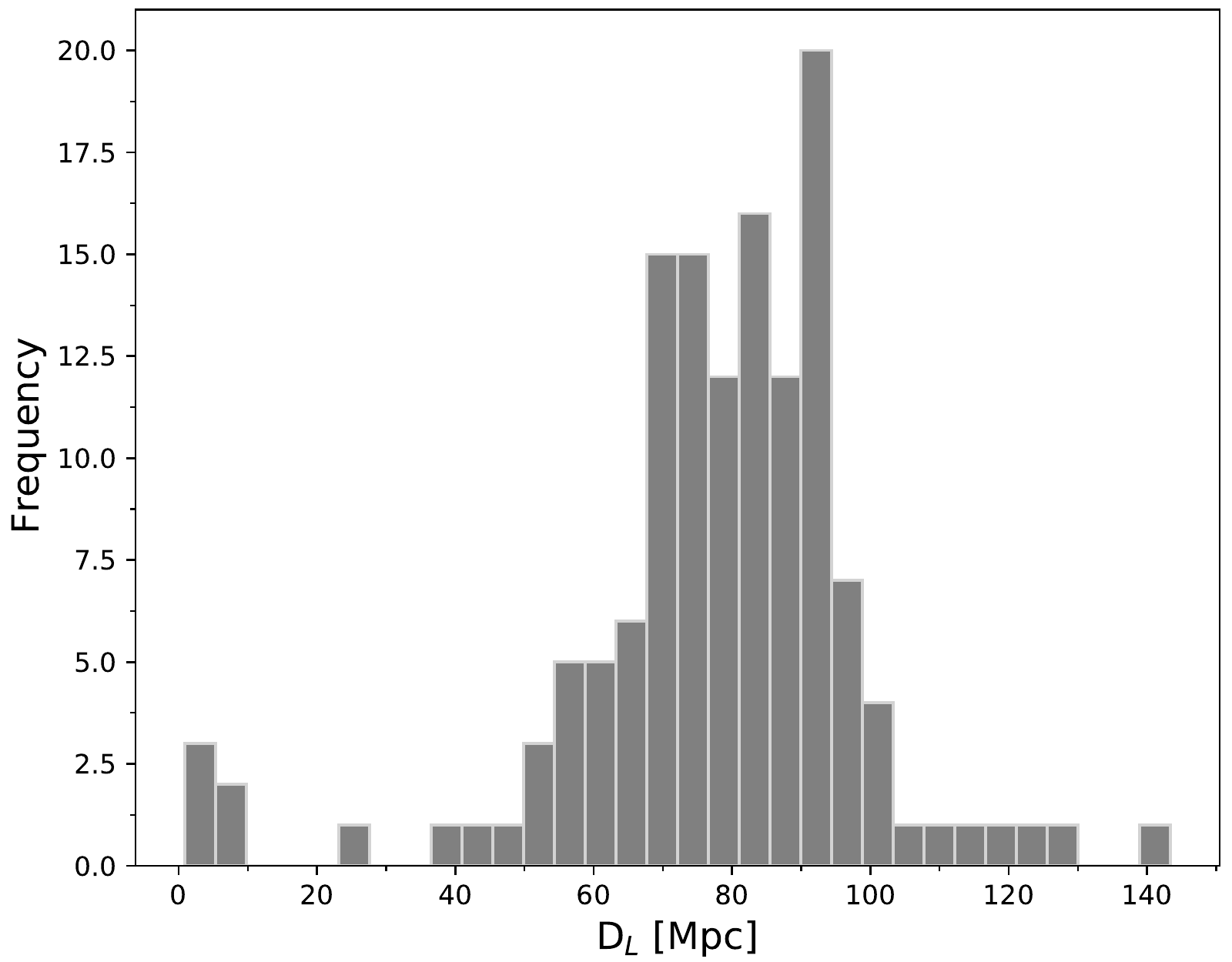}%
}\qquad

\subfloat{%
  \includegraphics[width=1\columnwidth]{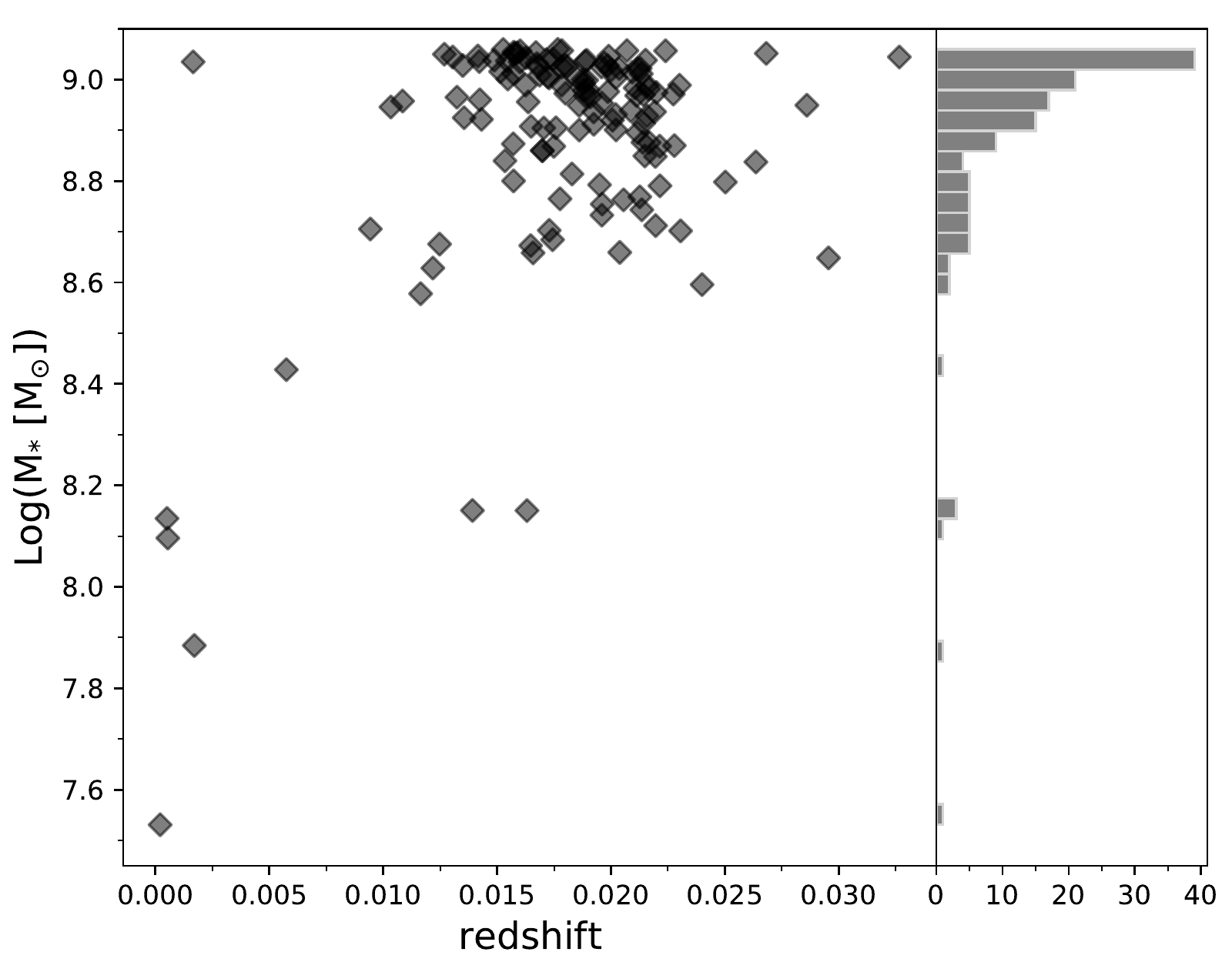}%
}
\caption{Upper panel: histogram representing the luminosity distance  distribution of the 136 galaxies that conform the current version of the MaNDala sample. Lower panel: relation between the redshift and M$_{*}$ for the galaxies in our sample, along with the distribution of M$_{*}$. Redshifts and M$_{*}$ are retrieved from the NSA Catalog.}
\label{fig:mass_z_histograms}
\end{figure}

In Figure \ref{fig:mass_z_histograms} we show the luminosity distance, $D_L$, and \ms{~} distributions in the upper and lower panels respectively, for this first version of the MaNDala sample. In the lower panel we also show the relation between redshift and \ms\ for our sample. We find that the MaNDala sample has the following limits in distance: 0.89 $<$ $D_{L}/{\rm Mpc}$ $<$ 143.37, with a mean of 77.48 Mpc. In the case of \ms, its range is: 7.53 $<$ log(\ms/\msun) $<$ 9.06, with a mean of Log(\ms/\msun)=8.89. This makes evident that our sample is biased towards galaxies that have \ms\ near the limit we imposed for the selection. The above suggests that our sample consist mainly of bright DGs, however, we leave the details about the sample characteristics to be presented in the Results Section (Sec. \ref{phot_characterization}).

\section{Photometric and Spectroscopic Analysis} \label{Analysis}

In this Section, we describe the photometric analysis of the MaNDala galaxy sample based on the DESI images (\S\S \ref{Analysis_Isophotes}), as well as our S\'ersic fits (\S\S \ref{Analysis_Profiles}). In \S\S \ref{Analysis_spectra}, we describe the spectral analysis performed to the MaNGA data.

\subsection{Isophotal Analysis} \label{Analysis_Isophotes}

We follow the iterative method of \citet{Jedrzejewski1987} to fit the isophotes of galaxies in the $g$, $r$ and $z$ band images from the DESI Legacy Imaging Surveys with a set of ellipses  using the IRAF \footnote{(Image Reduction and Analysis Facility) is distributed by the National Optical Astronomy Observatory, which is operated by the Association of Universities for Research in Astronomy, Inc., under cooperative agreement with the National Science Foundation.} task ELLIPSE. In our implementation, the ellipses are sampled along the semi-major axis of a galaxy in logarithmic intervals, most of the time starting from the intermediate to outer regions and decreasing the radius of each successive ellipse by a factor of 
$\sim0.9$. A trial and error procedure was used with different starting major-axis lengths to check the stability of the extracted parameters.

To estimate the center of a galaxy we proceed as follows. For regular-shaped galaxies we used the barycentric position of the light distribution in the central 5 pix $\times$ 5 pix region in the $r$-band images after applying IRAF image routines. For irregular-shaped galaxies with strong clumps and dusty regions, a careful masking of those clumpy regions was carried out and the center was estimated by applying the best-fit ellipses starting from the outer-most regions towards the central region, setting the center, position angle (PA), and ellipticity ($\epsilon$) as free parameters.

The $r$-band images were selected as the fiducial reference because of their relative lower sensitivity to dust extinction, high signal, and its relative good seeing. Once the center was estimated, it was fixed and the fitting started from the intermediate/outer regions of a galaxy while the position angle (PA) and ellipticity ($\epsilon$) were set as free parameters. Different values of the initial outer semi-major axis length were tried allowing us to check the consistency of the fitted ellipses and their quality flagging.  

A final step considers the extraction of the average isophotes in the $g$ and $z$ bands by using as a reference the already estimated $r$-band isophotal parameters. The above is to ensuring the extraction of a uniform profile and allowing for an estimate of color profiles from the combination of different bands.  
 
As stated in the DESI DR9 description,\footnote{\url{www.legacysurvey.org/dr9/description/}}
the pipeline removes a sky level that includes a sky pattern, an illumination correction, and a single, scaled fringe pattern. These corrections are intended to make the sky level in the processed images near zero, and to remove most pattern artifacts. In practically all cases in the MaNDala sample, the galaxy image is small enough that our retrieved frames contain portions of the sky unaffected by the galaxy, so the sky background corrections already implemented are adopted without any further correction. 

We also derived various image products from the reduced, calibrated images, namely, color index ($g-z$) maps and filter-enhanced images in the $r$-band optimized to enhance inner structures as well as low SB outer structures. These images were combined with the available RGB color images from the DESI legacy archives to generate image mosaics for each galaxy, very useful for the visual recognition of morphological details. 

\begin{figure}
\centering
  \includegraphics[width=1\columnwidth]{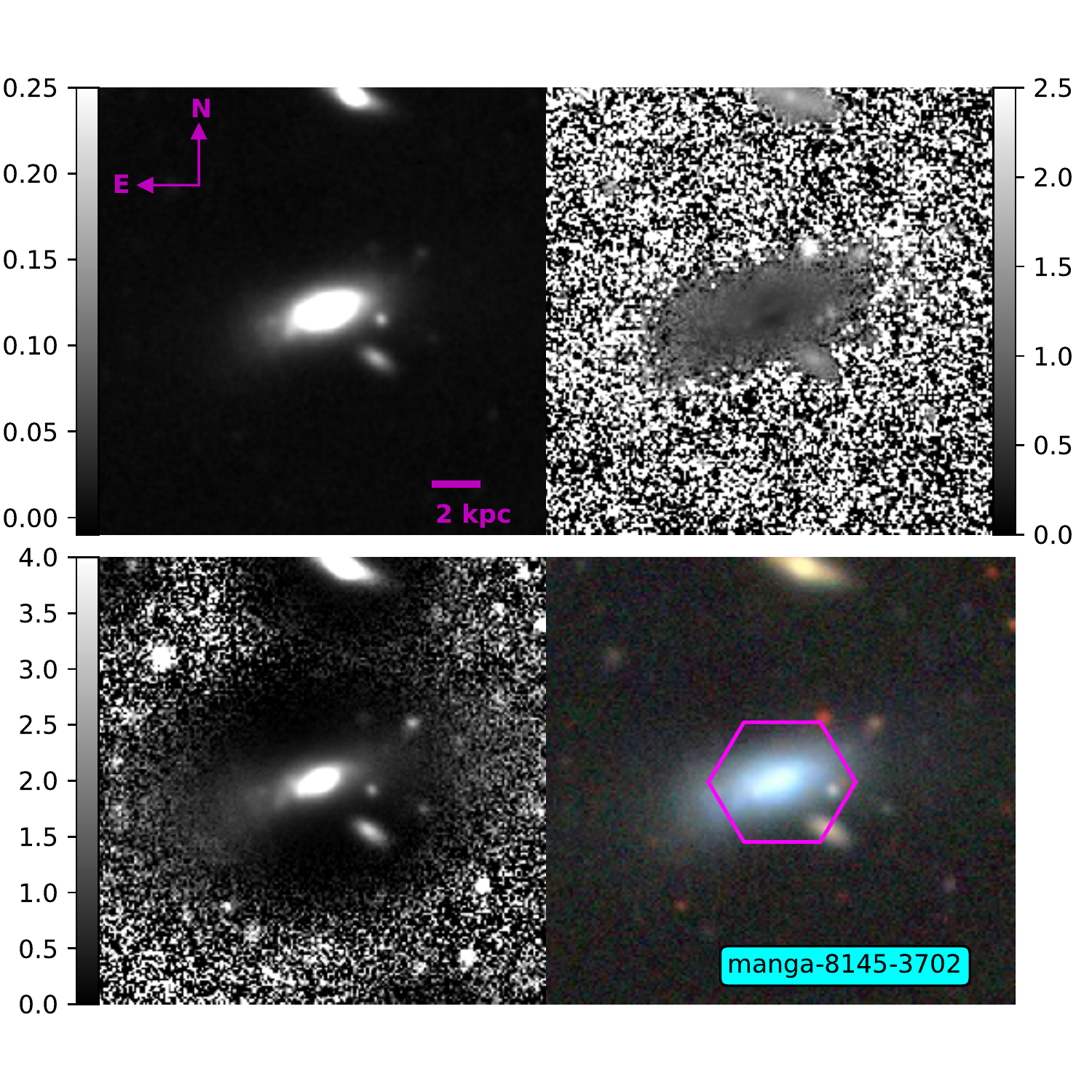}
\caption{Mosaic displaying the images derived from the photometric analysis for the galaxy MaNGA-8145-3702. Upper panels: $r$-band DESI image (left) and $g-z$ 2D color map (right). Lower panels: $g$-band DESI image with a Gaussian filter (left) and RGB DESI image. The hexagon in the RGB image represents the MaNGA field of view.}
\label{fig:manga-8145-3702-mosaic}
\end{figure}

In Figure \ref{fig:manga-8145-3702-mosaic} we show an example of the product images for the galaxy MaNGA-8145-3702, in the form of a mosaic. In the clock-wise direction and from the top left; an $r$-band DESI image, a $g-z$ 2D color map, an RGB DESI image and $g$-band DESI image post-processed with a Gaussian kernel ($\sigma$ = 15) and normalized to enhance low SB features over the background.

\subsection{Geometric Parameters and Surface Brightness Profiles}\label{Analysis_Profiles}

The ELLIPSE task in the IRAF STSDAS package estimates the intensity distribution along ellipses, which are expressed as a Fourier series:
\begin{equation}
I(\phi) = I_o + \Sigma a_n \sin(n\phi) + \Sigma b_n \cos(n\phi)
\end{equation}
where $\phi$ is the ellipse eccentric anomaly, $I_o$ is the mean intensity along the ellipse, while $a_n$ and $b_n$ are harmonic amplitudes.

The fitting started in the intermediate regions of galaxies going first to the outer regions up to a point where the mean counts are comparable to the $\sigma$ sky background. At this point the algorithm then goes back to the central regions and stops at the specified first central pixel.  

Foreground stars as well as apparently nearby galaxies and other image artifacts like diffracting patterns were carefully masked before ellipse fitting. However, in some galaxies the fitting could be distorted by mergers or the contaminating light from advanced galaxy interactions that our masking could not eliminate, causing $\epsilon$ and $PA$ to deviate to an arbitrary trend. In these cases we proceeded with the ellipse fitting either stopping near the edge of the galaxy or stopping farther out but taking note of these circumstances.  

Ellipse fittings proceeded by keeping the center fixed and allowing the $\epsilon$ and $PA$ parameters to vary in order to maximize the detection of inner structures like bars and other prominent features. Notice however that the fitting may be affected in the very central regions, due to algorithm indeterminacy in the inner-most 3-4 pixels, as described in the documentation of the ELLIPSE task in IRAF (see also \citep{Jedrzejewski1987}), or by seeing effects.

We propagate the errors on $I$ into errors on $\mu$ in magnitude units. The surface brightness profiles in the $g$ and $z$ bands are constrained to have the same geometric parameters as determined in the $r$ band. To build 1-D color profiles we proceeded by subtracting point-to-point the surface brightness of the $g$ band from that of the $z$ band, from the central regions up to where the SB profile of the $z$ band attains a 1$\sigma$ SB limit typically at $\sim 27.7$ mag arcsec$^{-2}$ according to our own estimates, based on the corresponding variance images.  

Finally, all SB profiles were inspected in order to ensure positive values. In cases of negative values at the end of their SB profiles,  those points were excluded. The presence of negative values was more frequently found in the $z$ band suggesting the influence of the background level at brighter levels compared to those in the $g$ and $r$ bands. A final cut of all the SB profiles is based on our own depth limits estimates for each galaxy in the three DESI photometric bands. 

The three surveys (DECaLS, BASS, and MzLS) use a three-pass strategy to tile the sky. This strategy is designed to account for the gaps between CCDs in the cameras, to ensure that the surveys reach the required depth, to remove particle events and other systematics, and to ensure photometric and image quality uniformity across the entire survey. For the Legacy Surveys, a post-processing catalog generation pipeline called legacypipe was created and The Legacy Surveys footprint is analyzed.\footnote{For the index of the Legacy Survey products, see \url{https://portal.nersc.gov/project/cosmo/data/legacysurvey/dr9/}}

Among the different images there contained, the label ``imag'' refers to files with the image pixels, the label ``invva'' refers to the surface-brightness uncertainties (inverse-variance) images, while the labels ``psfdepth'' and ``galdepth'' refer to estimates of the point-source or compact-galaxy detection levels. To estimate SB limits on an individual basis, we have retrieved the ``invvar'' images containing $1/\sigma^{2}$ for the pixels in the $grz$ bands and proceeded by reproducing our isophotal analysis (adopting the geometric $\epsilon$, PA and $R_{max}$ already obtained for the "image" files) on those images. Our ($5\sigma$) SB limits correspond to an isophotal annulus region around $R_{max}$ in each band. For 119 galaxies, their positions were close to the center of the retrieved inverse-variance maps. For the remaining galaxies their positions appeared off-centered so we proceeded with the SB limit estimates only after a more careful identification of each galaxy on these maps.
The mean values of these limits are: 27.69, 27.02 and 25.92 mag arcsec$^{-2}$ for the $g$, $r$ and $z$ bands, respectively. The individual SB limits for each galaxy in the $grz$ bands are retrievable directly from our website.\footnote{\url{https://mandalasample.wordpress.com/download/}} These limits are the ones adopted when performing Sersic fits to the SB profiles and other analyses described in the forthcoming sections.

\begin{figure*}
\centering
  \includegraphics[width=1\textwidth]{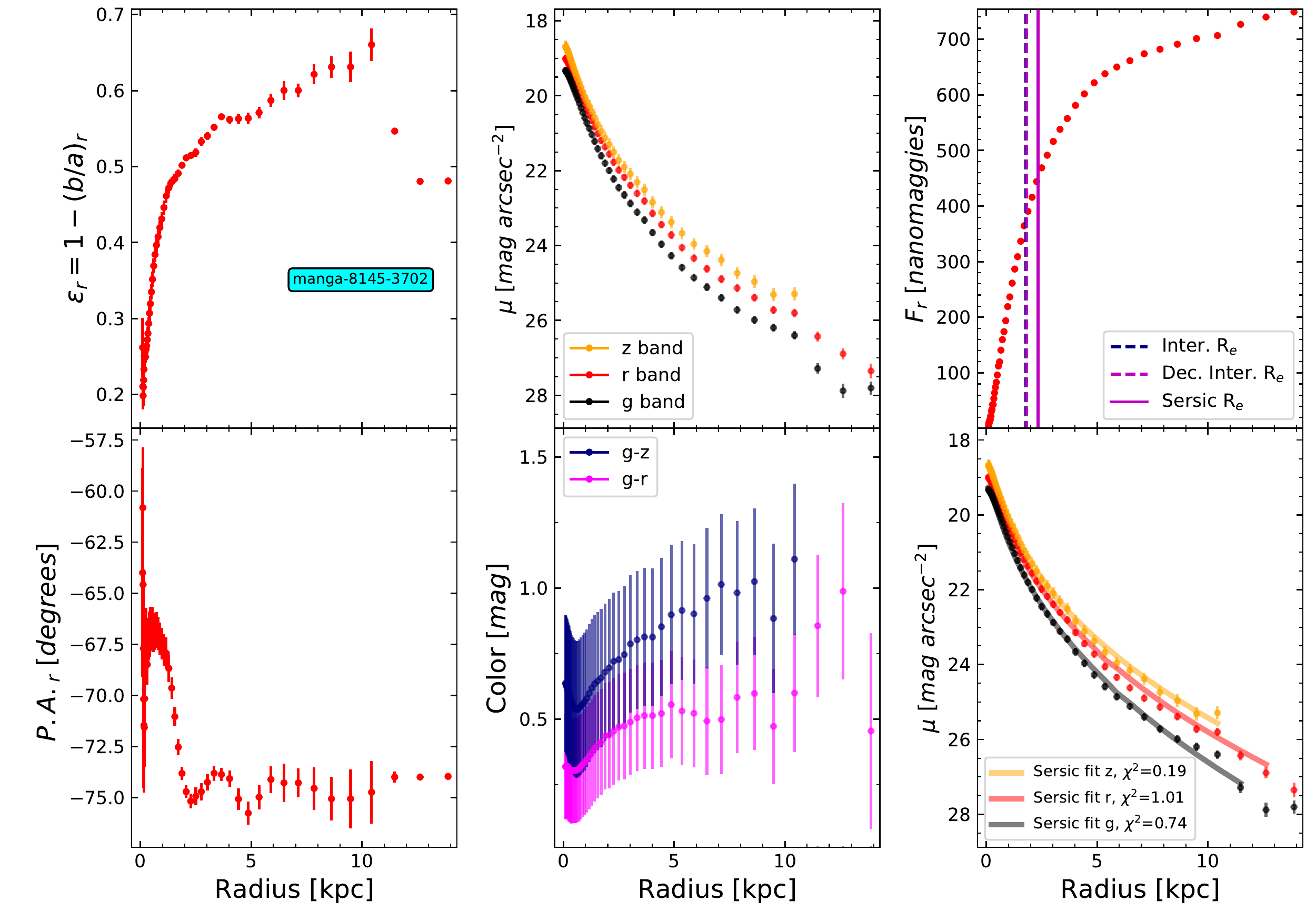}
\caption{Example of the radial profiles derived form the photometric analysis for the MaNGA-8145-3702 galaxy. Top panel from left to right: ellipticity ($\epsilon$) measured in the r-band, surface brightness profiles for the $z$, $r$ and $g$ bands (in yellow, red and black respectively) and the cumulative flux in the r band, along with three estimates of the $R_{e}$, from an interpolation of this curve, from the same interpolation but deconvolving with the PSF and from the Sersic fit in the $r$ band. Bottom panels from left to right: P.A measured in the r band, the $g-z$ and $g-r$ color profiles (in blue and magenta respectively) and Sersic fits to the surface brightness profiles for the three bands, displayed in the same color-code as the top-middle panel.}
\label{fig:manga-8145-3702-profiles}
\end{figure*}

In Figure \ref{fig:manga-8145-3702-profiles} we show an example of the collection of profiles that emerge from the photometric analysis for the galaxy MaNGA-8145-3702, which has already been presented in Figure \ref{fig:manga-8145-3702-mosaic}. In the clock-wise direction starting from the top left panel we show the following profiles: ellipticity ($\epsilon$) measured in the $r$-band, SB profiles for the $z, r,$ and $g$ bands (in yellow, red, and black respectively), the cumulative flux in the $r$ band, the Sersic fits to the SB profiles (to be explained in Section \ref{Results_Sersic}), the $g-z$ and $g-r$ color profiles (in blue and magenta respectively), and the P.A. measured in the $r$ band.

\subsubsection{Sersic Fit}\label{Results_Sersic}

Here we describe the fitting process to the SB profiles of the MaNDala galaxies. 

As it is often in the literature, we assume that the SB profiles of the galaxies are well describe by a \citet{Sersic_1963} function:
\begin{equation}
    I(R) = I_e \exp\left( -b_n  \left[
                    \left( \frac{R}{R_e} \right)^{1/n} - 1 
                    \right]\right),
                    \label{eq:sersic_profile}
\end{equation}      
where $R_e$ is effective radius, $n$ is the Sersic index, $I_{e}$ is the amplitude of the SB at $R_e$, and $b_n$ is such that
$\gamma(2n,b_n) = \Gamma(2n)/2$ (where $\Gamma$ and $\gamma$ are respectively the complete and incomplete gamma functions). In 
this paper we use the analytical approximation for $b_n$ reported in \citet{Ciotti_Bertin1999}, which we assume to be valid for $0.5\leq n\leq 10$. 
We note that in this paper the variable $R$ refers 
to the radius along the semi-major axis of the SB profiles. Therefore, the effective radius, $R_e$,
reported here will refer to the effective radius along 
the semi-major axis. 

The effects of seeing on the SB profiles of MaNDala galaxies are introduced by assuming that the PSF from the DESI images are well described by a \citet{Moffat1969} function with $\beta = 2.480$, $2.229$ and $1.999$ for the $g-$, $r-$ and $z-$bands respectively, (DESI help desk and Imaging Survey Experts, private communication). Thus, here we convolve the Sersic profile, Eq. (\ref{eq:sersic_profile}), by a Moffat PSF \citep[for a discussion see,][]{Trujillo+2001}; we will denote the above by $I_\text{conv}$. That is, assuming that the 1D profile is in elliptical coordinates $(x,y) = (\xi \cos \theta,\xi (1-\epsilon) \sin \theta)$ then the convolved SB along the semimajor axis, $\theta = 0$, is given by:
    \begin{equation}
        I_\text{conv}(\xi) = (1-\epsilon)
                        \int \int \text{PSF}(\xi',\theta',\xi)
                                    I(\xi')\xi'd\xi'd\theta'.
    \end{equation}

Here,  $\epsilon$ is the ellipticity, which for simplicity, we assume constant and equal to its average value for each galaxy. 

The methodology to determine the best fit parameters of Eq. (\ref{eq:sersic_profile}) for every galaxy in the MaNDala sample 
is as follows:
\begin{enumerate}
    \item As an initial guess, we use the $R_e$ from the photometric
    analysis described in a previous subsection and compute 
    $\mu_e(=-2.5\log I_e)$ from the observed SB profiles. Initially, 
    we assume that all galaxies have a S\'ersic index of
    $n=2.5$\footnote{This is a reasonable assumption since $n=2.5$ is half-way between disks and spheroids.}. 
    
    \item We sample the best fit parameters that minimize 
    the likelihood function $L\propto e^{\chi^2/2}$ by using 
    the Markov Chain Monte Carlo method \citep[described in][]{Rodriguez-Puebla+2013} and the initial values described in the
    previous item. We run 10 chains consisting of $10^4$ elements each 
    and $\chi^2$ is given by:
    \begin{equation}
        \chi^2 = \sum_{i=1}^{N_{\rm bin}} \left(
                        \frac{ \mu_{i,\rm model} -  \mu_{i,\rm obs} }{\sigma_{i,\rm obs}}
                        \right)^2,
                        \label{eq:chi2}
    \end{equation}
    where $N_{\rm bin}$ is the number of radial bins in the observed SB profiles $\mu_{i,\rm obs}$ of each galaxy and $\sigma_{i,\rm obs}$ as its corresponding error, and $\mu_{i,\rm model}$ is the SB profiles given by Eq. (\ref{eq:sersic_profile}). 
    As a result, we find the best fit parameters to $\mu_e$, $n$ 
    and $R_e$ that minimize Eq. (\ref{eq:sersic_profile}).
    
    \item Next, we use the best fitting parameters and the 
    covariance matrix constrained above
    as the initial guess for finding the best fit parameters of 
    the convolved S\'ersic profile, $I_{\rm conv}(R)$. We do so by replacing
    in Eq. (\ref{eq:chi2}) $\mu_{\rm model}$ by $\mu_{\rm conv-model}$.
    Here we sample 3 chains consisting of 500 elements each.\footnote{The number of elements and chains is reduced because this is a computational intensive calculation. However, we find that the above setting is enough to sample the space parameter due to the optimization in the priors.} 
\end{enumerate}

The best fitting parameters described here are fitted to each band independently. That is, we do not make any assumption on the wavelength dependence of the S\'ersic profile parameters. 
The bottom right  panel of Figure \ref{fig:manga-8145-3702-profiles} shows an example to the best fit S\'ersic profiles to the observed SB profiles of MaNGA-8145-3702 galaxy. The inset in the same panel shows the reduced $\chi^2$ defined
as $\chi^2/({\rm d.o.f})$, where ${\rm d.o.f} = N_{\rm bin} - 3$. 



\subsection{Spectroscopy analysis}
\label{Analysis_spectra}


For this work, we use the data products provided by the 3.1.1 version of the SDSS-IV Pipe3D Value Added Catalog\footnote{\url{https://www.sdss.org/dr17/manga/manga-data/manga-pipe3d-value-added-catalog/}} \citep[VAC; Sanchez et. al submitted,][]{,sanchez18a}. For our purposes, we have homogenized these data products to be consistent with the cosmological model adopted by us (Pipe3D VAC data products use: $H_0$ = 71 km/s/Mpc, $\Omega_{M}$= 0.27, $\Omega_{\Lambda}$ = 0.73).

We briefly summarize how {\sc pyFIT3D} works \citep[for more details we refer the reader to][]{Sanchez16a,Sanchez16b,Lacerda+2022}: A spatial binning is first performed in order to reach a S/N of 50 per bin across the entire field of view (FoV). A non-parametric stellar population synthesis (SPS) analysis is then applied to the co-added spectra within each spatial bin. The SPS analysis fits the continuum to a compose set of simple stellar populations (SSPs) of 39 ages, linearly spaced for ages of $<0.02$ Gyr and logarithmically spaced at larger ages,  and 7 metallicities: $Z_\star$= 0001, 0005,0.0080, 0.0170, 0020, 0.0300, 0.0400.
The SSPs were generated with an updated version of the {\sc Bruzual \& Charlot 03} SPS models (Bruzual et al. in prep., Sanchez et al., submitted), using the \verb|MaSTAR| stellar library \citep{MaSTARS} and a \citet{Salpeter+55} IMF. We refer to this set as \verb|sLOG|. The use of \verb|sLOG| improves the SPS for the MaNDala galaxies mainly by extending the metallicity range to lower values. Finally, the \citet{Cardelli+1989} extinction law is used in the calculation of the dust attenuation.%

{\sc pyPipe3D} re-scales the stellar population models for each spaxel  within each spatial bin to the continuum flux intensity in the corresponding spaxel, and generate a set of spatially resolved maps of the SPS properties. In this paper, we used the information from the spatially resolved maps of the luminosity- and mass-weighted stellar ages in each spaxel that are calculated as logarithmic averages:
\begin{equation}
\log(age_{mw})=\sum_j^{n_{ssp}}\log(age_{ssp,j})m_{ssp,j}/\sum_j^{n_{ssp}}m_{ssp,j},
\label{eq_ageMW}
\end{equation}
\begin{equation}
\log(age_{lw})=\sum_j^{n_{ssp}}\log(age_{ssp,j})L_{ssp,j}/\sum_j^{n_{ssp}}L_{ssp,j},
\label{eq_ageLW}
\end{equation}
where the $j$ SSP is characterized by the age $age_{ssp,j}$, luminosity $L_{ssp,j}$, and mass $m_{ssp,j}$;
$n_{ssp}$ is the total number of SSPs. 
The mass- and luminosity-weighted stellar metallicities are calculated in the same way.
We use the mass- and luminosity-weighted age and metallicity maps to derive the respective stellar ages and metallicities up to $R_{e}$ and the entire FoV, in the $r$-band. To estimate the aperture within $R_{e}$, we use the $P.A.$ and ellipticity values reported in this work, while the FoV aperture corresponds to the IFU bundle area of the given target. 


\subsubsection{Ionized gas emission lines}

{\sc pyPipe3D} subtracts the fitted stellar population models from the original data cube to create a cube comprising only the ionised gas emission lines (and the noise). Individual nebular emission line fluxes are then calculated segment by segment using a weighted momentum analysis based on the kinematics of {\sc H}$\alpha$ \citep[Sanchez et al., in prep,][]{Lacerda+2022}. From the obtained emission line maps, we integrate the ionized gas line fluxes within the two apertures mentioned above, $R_{e}$ and the FoV of each galaxy. To avoid contamination of any nearby star within our apertures, we used a set of star masks derived from the GAIA star positions. These star masks are described in Sanchez et al., in prep. 
The nebular emission lines used in this work are: {\sc H}$\beta$, {\sc H}$\alpha$, {[O\sc iii]}$\lambda5007$, {[N\sc II]}$\lambda6584$, {[S\sc II]}$\lambda6716$, {[S\sc II]}$\lambda6731$, and {[O\sc I]}$\lambda6300$. 
We estimate the total equivalent widths (EW) of {\sc H}$\alpha$ within a given aperture by dividing the total emission line flux by the integrated stellar continuum within the same aperture.

\section{Results}\label{Results}

\subsection{Basic Morphological Classification}\label{Results_Morphology}

sThe MaNGA Visual Morphology Catalog\footnote{\url{https://www.sdss.org/dr17/data_access/value-added-catalogs/}} \citep{VazquezMata2022} provides a classification in terms of the Hubble Sequence for MaNGA galaxies. Since the MaNDala sample is a subset of the complete MaNGA sample, that classification has been inherited here. Although other classification schemes are more appropriate for dwarf galaxies \citep[see for example:][]{Kim+2014} such classification is not available at this time for the MaNDala sample, then we preferred to keep the Hubble classification only for descriptive purposes. On line with this, we describe the classification in terms of two broad Hubble type groups: Early types, comprising E and Sa types, and Late types, comprising types equal or later than Sab.


We are aware that the classification scheme chosen is tentative and that a more detailed classification, using the specific morphological types for dwarfs is required. However, for the purpose of this work, it is enough to have this basic morphological information. We were able to provide a classification for 135 galaxies from our sample, with one remaining, due to its intrinsic faintness. Our results show that 25 belong to the Early group ($\sim$19\%) and 110 to the Late ($\sim$81\%). As expected the vast majority of the galaxies in our sample of bright dwarfs fall within the Late morphological group, which contains mostly vary late to Irregular types.

\subsection{Environment}\label{Results_Environment}

A simple way to characterize the environment of galaxies is dividing them between central and satellites. For this purpose we make use of the information given in the SDSS-IV Galaxy Environment for MaNGA VAC (GEMA),\footnote{\url{https://www.sdss.org/dr17/data_access/value-added-catalogs/?vac_id=gema-vac:-galaxy-environment-for-manga-value-added-catalog}} which utilizes the methodology described in \citet{Yang2007} to identify galaxy groups composed of a central galaxy and its satellites. 121 MaNDala galaxies are in this catalog, for which we identify 86 centrals and 35 satellites. 82 of the central galaxies are the most massive within their groups, and 76 of them belong to a group of only one galaxy. Another way to characterize the environment of DGs is through the distance to its nearest luminous neighbor, $D_{host}$. To achieve this, we use the determinations of this parameter given in \citet{Geha2012}. Their environment criterion is defined in the following way: a galaxy beyond $D_{host} > 1.5$ Mpc (within 1000 km s$^{-1}$ in redshift) of a luminous host galaxy is defined to be in the field, which implies that it is relatively isolated, otherwise it is considered as not isolated. For all of the galaxies in our sample we have this information, however we relax the environment criterion while using a threshold value of 1 Mpc for $D_{host}$ for which we identify 63 ($\sim$46\%) to be in the field and 73 ($\sim$54\%) in denser environments. The main reason for changing the value of $D_{host}$ is because clusters, of halo mass $M_{\rm vir}\sim10^{14}\msun$, are  larger than $\sim$1 Mpc. This means that the value adopted by \citet{Geha2012} could be larger than cluster environment, in other words, their field galaxies would be biased to very low density environments.

\begin{figure}
\centering

\subfloat{%
  \includegraphics[width=1\columnwidth]{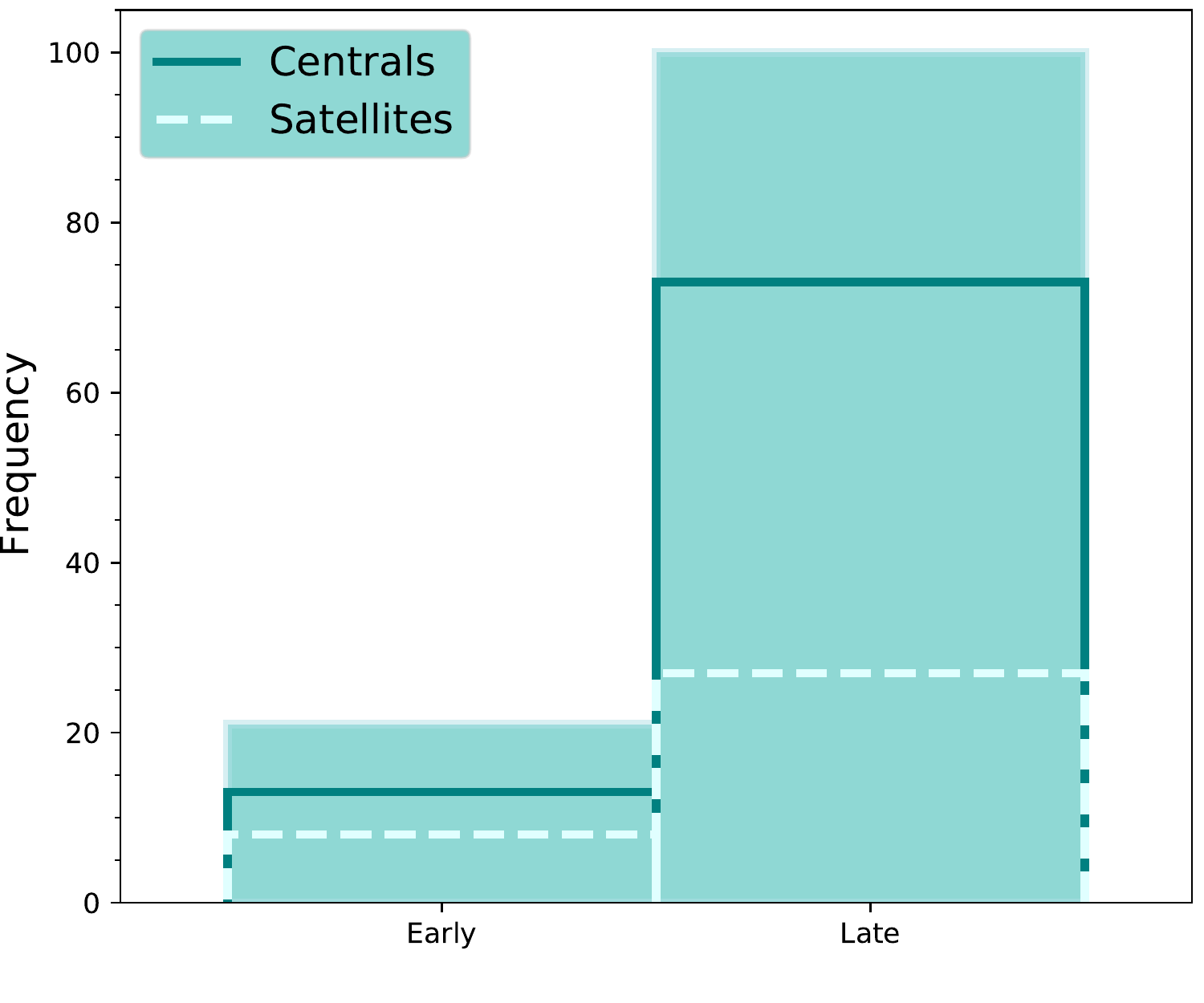}%
}\qquad

\subfloat{%
  \includegraphics[width=1\columnwidth]{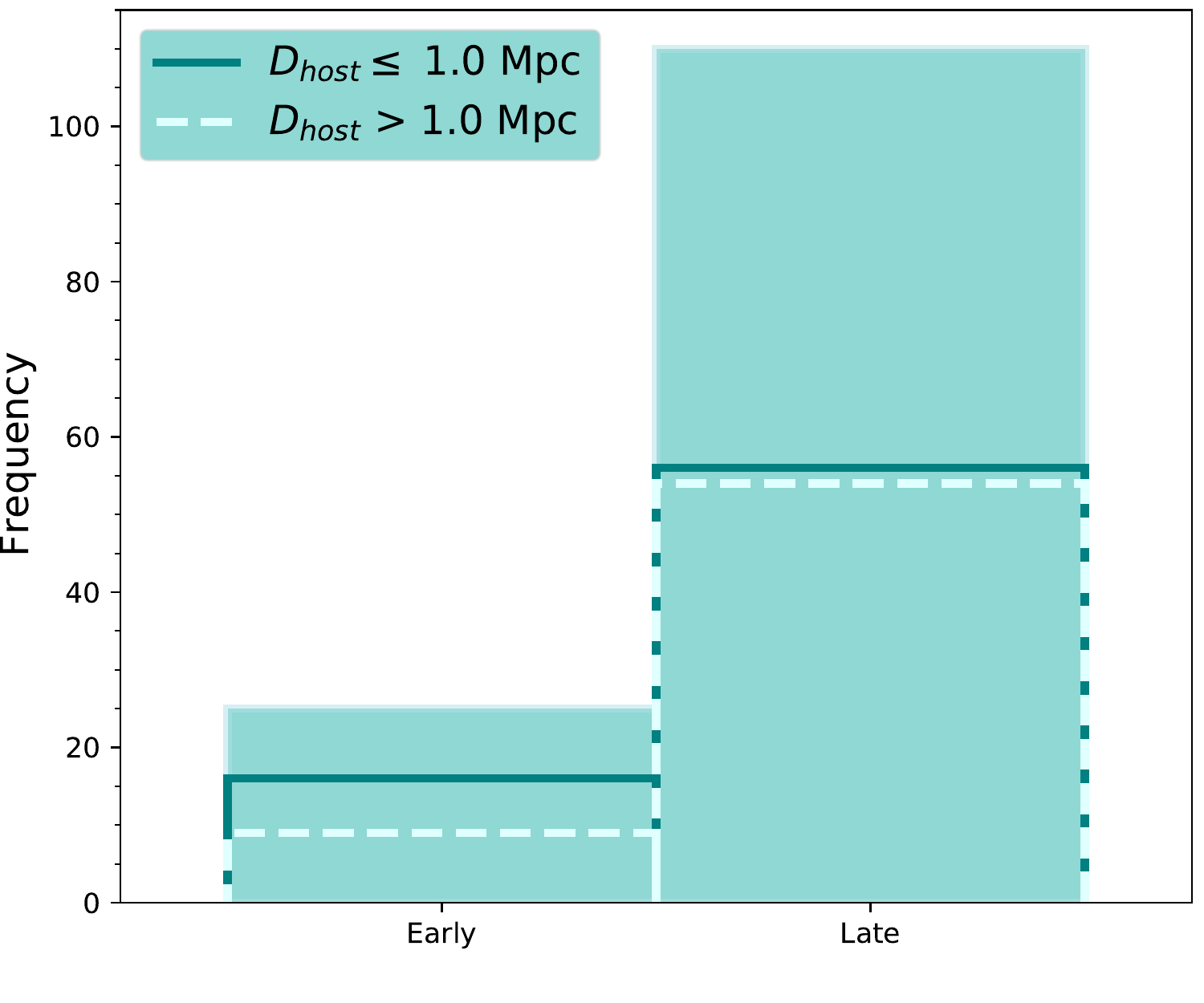}%
}
\caption{Upper panel: filled green bars show the histogram of the two morphological types for the 121 MaNDala galaxies with central/satellite information available in the GEMA Catalog. Empty bars show the histograms for the same morphological types only for central and satellite galaxies in solid and dashed lines respectively. Lower panel: filled green bars show the histogram of the two morphological types for the 135 MaNDala galaxies with information from \citet{Geha2012} on the distance, $D_{host}$, to the nearest luminous galaxy. Solid and dashed line bars show the histograms for field ($D_{host}>1$ Mpc, relatively isolated) galaxies and those in more dense environments ($D_{host}\le1$ Mpc).}
\label{fig:environment_histograms}
\end{figure}

In the upper panel of Figure \ref{fig:environment_histograms} we present a histogram of the morphological types for the 121 galaxies in our sample found in the GEMA Catalog. The filled green bars represent the complete set of 121 galaxies separated into Early and Late types. The solid line bars correspond to those galaxies identified as centrals, while the dashed line bars correspond to the satellites. In the lower panel we show a similar histogram, in which the filled bars represent the 135 galaxies for which we have morphological and environmental information from \citet{Geha2012}, with dashed and solid line bars corresponding to field and denser environments, respectively. Central dwarfs dominate for both late and early types. On the other hand, early type dwarfs in our sample tend to be in not isolated environments, while the late types tend to be in equal proportions in terms of the $D_{host}$ parameter threshold.

\subsection{Photometric Results}\label{Results_Photometry}

In this Section we present several results derived from the analysis performed with the photometric data, already described in the previous Section (\ref{Analysis_Isophotes}, \ref{Analysis_Profiles}), and how they compare to the public results given in the NSA Catalog. The images and profiles for all the galaxies in our sample, used to derive the results in this section are available as part of the SDSS-IV MaNDala VAC (for details, we refer to Appendix \ref{AppendixVAC}).

\subsubsection{Comparison with NSA}\label{Results_NSACompare}

 To ensure that our photometric analysis is consistent with previous results, here we present a series of comparisons between our results and those available in the NSA Catalog. We recall that our analysis is based on the photometry from the DESI Legacy Imaging Surveys, while in the NSA Catalog, the shallower SDSS photometry has been used.
 

\begin{figure}
\centering
\subfloat{%
  \includegraphics[width=1\columnwidth]{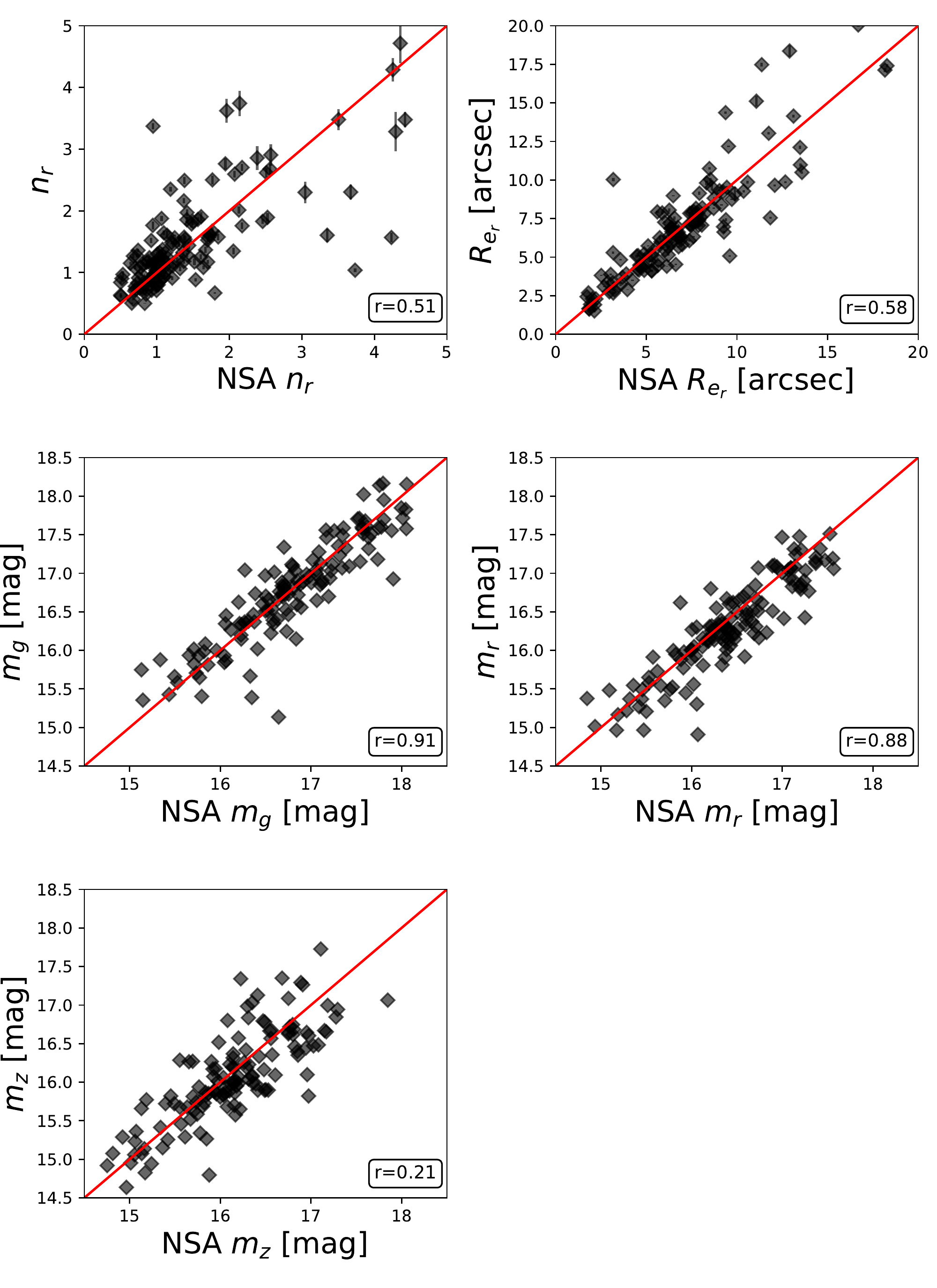}%
}
\caption{Comparisons of the principal results from our Sersic fit against those previously presented in the NSA Catalog. From the top to the bottom panels we show the following comparisons: Sersic index and effective radius in the $r$ band, and apparent magnitudes in $g, r, z$ bands. The red line shows the one-to-one relation. In all panels the Pearson correlation coefficient is shown.}
\label{fig:nsa_fit_compare}
\end{figure}

In the upper panels of Figure \ref{fig:nsa_fit_compare} we show the comparisons of the S\'ersic index and $R_{e}$\footnote{Sersic\_TH50 column reported in their public catalog.} derived from our photometric analysis and S\'ersic fit in the $r$-band against those given in the NSA Catalog\footnote{The Sersic\_TH50 column in their public catalog}. In both plots we see a good agreement with the NSA results. However, in both cases, when reaching the largest values, the dispersion in the points becomes larger. In the case of the index $n$ it is noticeable that for the largest values reported by the NSA ($n_{r}$ $>$ 2), we find smaller values, mostly between 1 and 2, which is more consistent for smaller and LTG galaxies. The opposite occurs in the case of R$_{e}$, for which in general, our estimations seem to be larger than those reported in the NSA. These both panels exhibit some outliers. We checked if the large differences between the Sersic indexes and $R_{e}$ derived by us and those reported in the NSA Catalogue for those objects may be due to bad fits of the SB profiles, however the majority of them have $\chi^{2}$ values below 3 for both plots. A possible source of the differences could be the fact that the NSA fits use an extrapolation while ours use only observed points with the SB profiles. 

In the two bottom panels of Figure \ref{fig:nsa_fit_compare} we show the comparisons between our estimations of the apparent magnitudes in the $g,r,$ and $z$ bands and those from the NSA Catalog. The NSA apparent magnitudes were derived using the absolute ones retrieved directly from their catalog. In all panels we also include the Pearson correlation coefficient (r). As in the previous plots, in general there is a good agreement, particularly in the $g$ and $r$ bands, while for the $z$ band the dispersion seems larger, which is also visible in its low r value.

Also recently \citep{Arora2021} performed a photometric analysis for 4585 MaNGA galaxies using DESI data. For the 71 galaxies that we have in common with their sample we performed comparisons for some global properties provided by them such as apparent magnitudes, $R_{e}$ and $\mu_{eff}$ and our results are consistent with theirs (Pearson coefficients of $\sim$0.9 for the first two and $\sim$0.8 for the last one). It is important to mention that the results given by \citep{Arora2021} are not based on S\'ersic fits, instead they are taken from the photometric analysis directly, meaning that these comparisons are not direct.

\subsubsection{Radial Surface Brightness and Color Profiles} \label{SB_color_profiles}

\begin{figure*}
\centering
\gridline{\fig{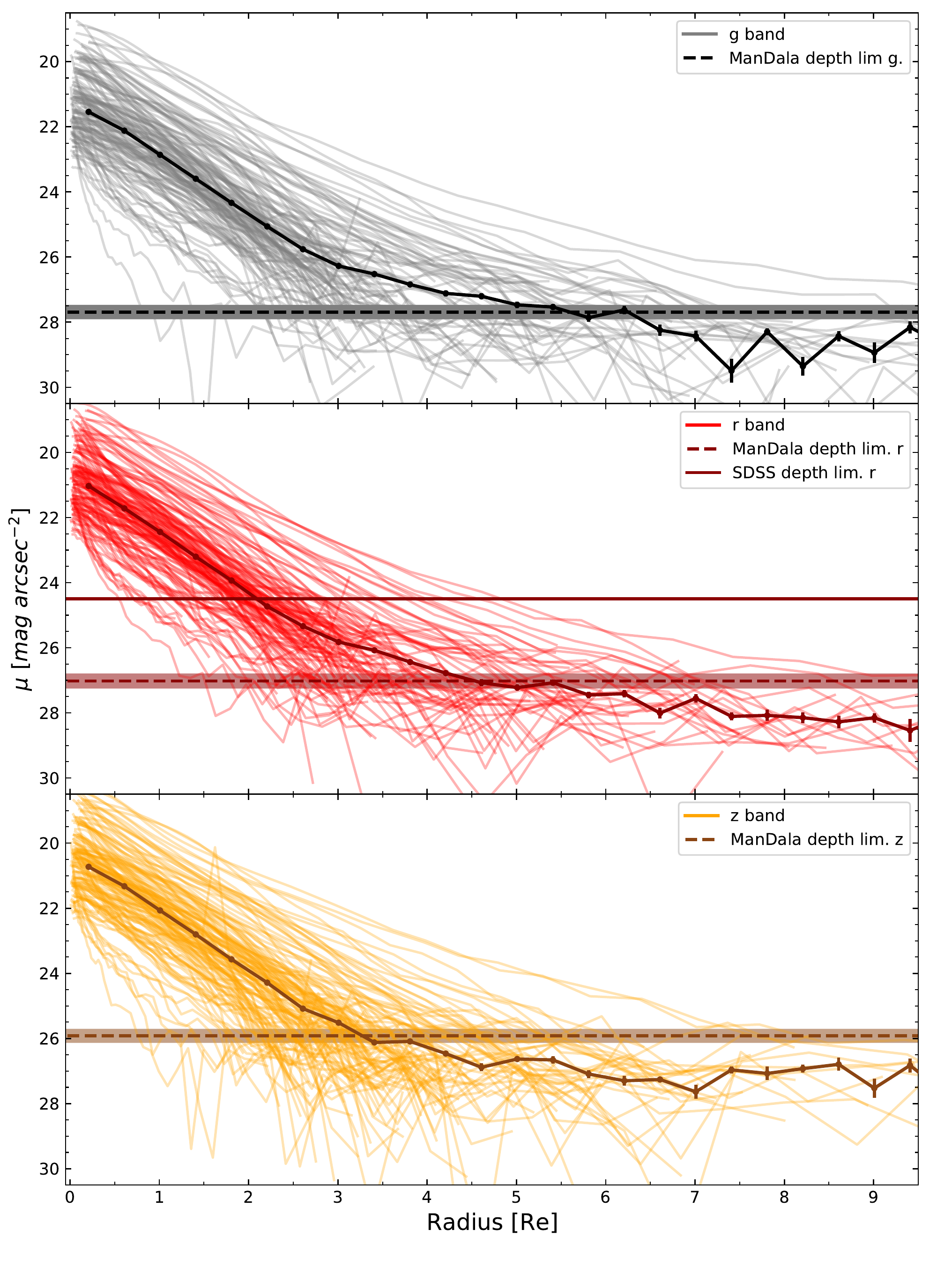}{0.5\textwidth}{}
          \fig{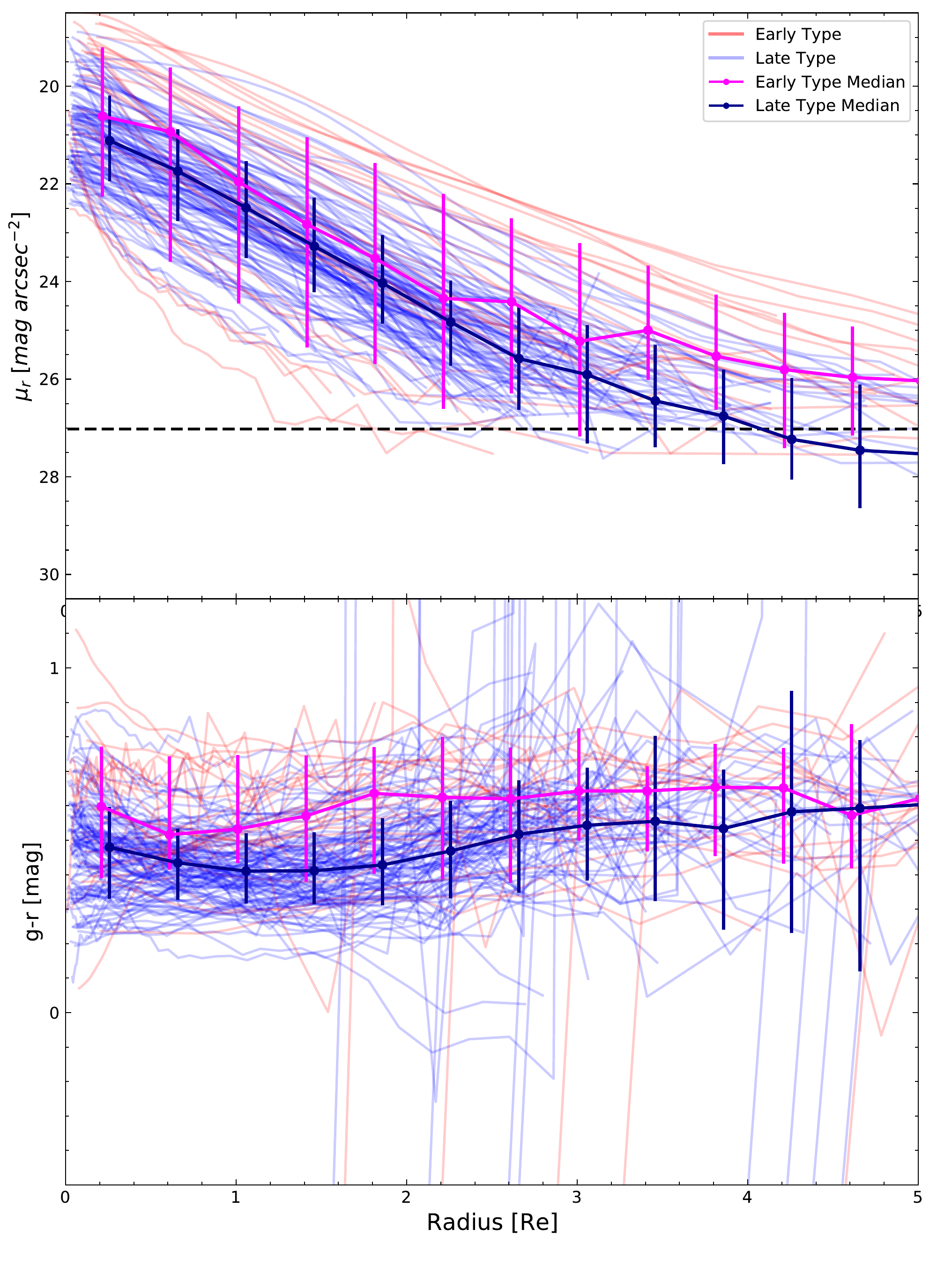}{0.5\textwidth}{}
        }
\caption{Left panels: from the uppermost to the bottom, the individual surface brightness profiles for all the MaNDala galaxies, in the $g$, $r$ and $z$ bands respectively. On top of them an average profile is shown, calculated in radial bins of 0.4 $R_{e}$. The dashed lines represent the mean depth limit values for the galaxies in the sample for each band, while 1$\sigma$ of those values are shown by the shaded areas around them. Right panels: in the top is shown the same surface brightness profiles for all the MaNDala galaxies in the $r$ band as in the right panel, with its depth limit in the dashed line, but color coded in terms of morphology (blue for late types and red for early types). In the bottom $g-r$ profiles for each MaNDala galaxy are shown, with the same color code as in the top. On top of both panels, the median profiles for early and late type galaxies are plotted in magenta and dark blue respectively, error bars represent the 16th and 84th percentiles. The surface brightness and color profiles shown in this figure are cut according the individual depth limit for each galaxy. \label{fig:SBandColorProf_CompareBands}}
\end{figure*}

In the left panels of Figure \ref{fig:SBandColorProf_CompareBands} we show the individual SB profiles for all the galaxies in our sample in the $g$, $r$ and $z$ bands (for each band, the radii are normalized to their corresponding $R_{e}$ derived from the Sersic fit). On top of them an average profile is shown for each band. In dashed lines and shaded areas, the mean depth limits for each band of the MaNDala sample and the standard deviations are shown respectively (see Section \ref{Analysis_Profiles} for their derivation details). For comparison, the red solid line shows the SDSS depth limit in the $r$ band at 24.5 mag arcsec$^{-2}$ \citep[see for example:][]{Strauss+2002}, which emphasize the difference in photometric depth achieved by DESI and SDSS image data. These plots give an idea of the difference between the photometric depths of the DESI data in the three bands. We can see that the flattening of the average profile rise to brighter SB values when moving from the $g$ to the $z$ band. This is relevant as the flattening of these profiles is related to the radius of the galaxies in which the brightness of the sky is starting to dominate in the data. For the $g$ band we are able to obtain the best SB profiles, as in average, they arrive out to $\sim 5 R_{e}$ before the profiles start to flatten. The flattening moves to inner regions of the galaxies when moving to redder bands, as it occurs at $\sim 4.5 R_{e}$ and $\sim 3.5 R_{e}$ in the $g$ and $z$ bands respectively (note that, as shown in Figure \ref{fig:sample_histograms} below, the measured effective radii are roughly similar in the three bands). It is also important to mention that NSA SB profiles for these galaxies can extend to larger radii, beyond the photometric depth limits of the SDSS images (see middle left panel of Figure \ref{fig:SBandColorProf_CompareBands}). In contrast, the profiles from our analysis naturally extend beyond the SDSS depths up to the DESI limits, showing the  shape of the profile obtained at those radii. 

In order to make evident any systematic difference provided by the morphology, in the top right panel of Figure \ref{fig:SBandColorProf_CompareBands} we reproduce the profiles in the $r$ band of the top left panel of the same Figure, but color coded according to the morphology: red and blue for the early and late types. In a second panel we also show the individual color $g-r$ profiles for our sample, color coded in the same way as in the previous panel. 
On top of both panels a median profile is shown in magenta for the early type galaxies and in dark blue for the late ones. These general profiles are derived binning the early and late types distributions with a bin size of 0.4 in units of $R_{e}$ and calculating the median value for each one. The error bars correspond to the 16th-84th percentiles of the distributions within each bin.

According to the right panel of Figure \ref{fig:SBandColorProf_CompareBands}, the $r$-band SB profiles of the dwarf LTGs are less scattered than those of the dwarf ETGs (note that the distribution of magnitudes or masses of both subsamples are roughly similar, and since LTGs are much more numerous than ETGs, the greater dispersion of the latter with respect to the former is hardly an effect of sample size.). On average, the ETGs have slightly higher SB at all radii than the LTGs. As for the $g-r$ color gradients, they fluctuate with radius but around a fixed value, that is, they tend to be flat or slightly positive, both for dwarf LTGs and ETGs. As expected, the latter are redder than the former.

\subsubsection{Photometric Characterization of the Sample} \label{phot_characterization}

Making use of our photometric results we can now perform a basic characterization of the MaNDala sample, such as identifying its limits, and even testing if all the objects are consistent with the expected behavior for DGs.




\begin{figure}
\centering
\subfloat{%
  \includegraphics[width=1\columnwidth]{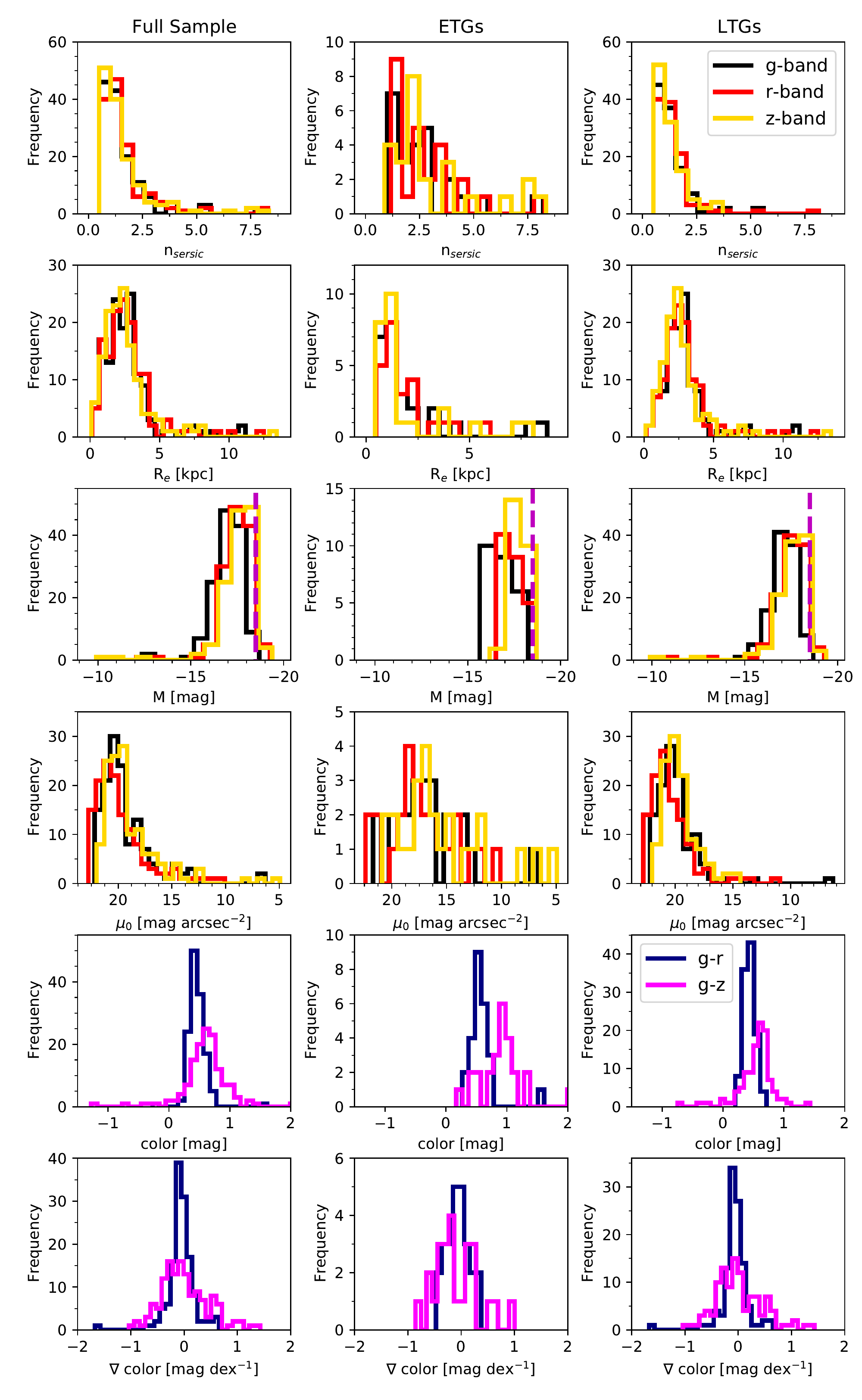}%
}
\caption{Set of histograms that characterize the full MaNDala sample (left column), the early type galaxies (middle columns) and the late types (right column). From the upper most to the bottom, the panels show the distributions of the following properties: Sersic index, effective radius, apparent magnitude, absolute magnitude, central surface brightness, color measured at the center and color gradients measured between 0.1 and 1 $R_{e}$ . The color code is the same used to present the surface brightness and color profiles in Figure \ref{fig:manga-8145-3702-profiles}. The magenta dashed line in the absolute magnitude panel marks the M$_{g}$ = -18.5 mag limit imposed to select this sample, used with the NSA Catalog photometric data.}
\label{fig:sample_histograms}
\end{figure}

Histograms of the S\'ersic index, effective radius, apparent and absolute magnitudes, central SB, central color and color gradients for the three DESI photometric bands used in this work for our sample, are shown in Figure \ref{fig:sample_histograms}. Color gradients are defined as:

    \begin{equation}
       \nabla (\lambda_{1}-\lambda_{2}) = \frac{\Delta(\lambda_{1}-\lambda_{2})(R)}{\Delta(logR)}.
                        \label{eq:color_gradient}
    \end{equation}
We evaluate $\nabla(\lambda_{1}-\lambda_{2})$ at 0.1 and 1 $R_{e}$. All of these properties are results of the S\'ersic fits explained in Section \ref{Results_Sersic}. These distributions show the natural limits of the sample in brightness, which are summarized in Table \ref{table_phot}, along with other relevant characteristics. In general, we can see that the MaNDala sample is indeed biased towards bright DGs, as suggested by the M$_{*}$ distribution shown in Figure \ref{fig:mass_z_histograms}. Moreover, the dashed magenta line in the absolute magnitude histogram represents the limit imposed to select the galaxies in our sample (NSA M$_{g}$ $>$ -18.5). It is noticeable that a few of our galaxies surpass this limit while using the results derived with the DESI photometry and with our S\'ersic fits, which as shown in the previous section, some galaxies tend to give brighter magnitudes when compared with those available in the NSA Catalog.

From Figure \ref{fig:sample_histograms}, we see that the S\'ersic index of dwarf ETGs tends to be higher and with a greater spread than for LTGs (there is also a small trend of lower $n$ values as the band is redder, specially for ETGs). As reported in Table \ref{table_phot}, the medians of $n_r$ for ETGs and LTGs are 2.50 and 1.15, and the 1-$\sigma$ scatters are 1.60 and 0.94, respectively. A similar trend is observed for $\mu_0$: ETGs tend to have higher central SB's and a greater spread than LTGs (median $\mu_{0,r}$ of 16.98 and 20.23 mag arcsec$^{-2}$, respectively. Regarding the sizes, $R_e$ depends little on the photometric band and tends to be smaller for the ETGs than for the LTGs (note that the absolute magnitude distributions are similar for both groups): the median $R_{e,r}$ are 1.33 and 2.58 kpc, respectively. In general and as expected, the dwarf ETGs are significantly more compact than the dwarf LTGs; the ratios of the median $r$-band luminosity to the median effective radius for ETGs is $\approx 1.7$ times higher than for LTGs. As for the colors, ETGs are redder on average than LTGs (medians of $g-r$ of 0.57 and 0.44 mag, respectively) but with a broader distribution. Finally, the $g-r$ color gradients of the MaNDala galaxies oscillate around 0, meaning that the gradients are nearly flat. 

\begin{table}
\footnotesize
\centering
    \begin{tabular}{ l c c c c c }
    \hline
    \hline
  & Max & Min & Mean & Median & $\sigma$ \\
\hline
\textbf{Full sample} & & & & & \\
n$_{r}$ & 8.30 & 0.50 & 1.64 & 1.25 & 1.22 \\
$R_{e_{r}}$ [kpc] & 12.49 & 0.15 & 2.71 & 2.45 & 1.76 \\
m$_{r}$ [mag] & 17.61 & 12.28 & 16.18 & 16.27 & 0.87 \\ 
M$_{r}$ [mag] & -10.63 & -19.27 & -17.38 & -17.48 & 1.05 \\ 
$\mu_{0_{r}}$ [mag arcsec$^{-2}$] & 22.20 & 6.30 & 19.29 & 19.98 & 2.58 \\
g-r [mag] & 1.61 & -0.27 & 0.46 & 0.44 & 0.16 \\
$\nabla (g-r)$ [mag dex$^{-1}$] & 0.66 & -1.67 & -0.05 & -0.06 & 0.25 \\
\textbf{Early Types} & & & & & \\
n$_{r}$ & 8.30 & 1.18 & 2.77 & 2.50 & 1.60 \\
$R_{e_{r}}$ [kpc] & 6.02 & 0.48 & 1.93 & 1.33 & 1.46 \\
m$_{r}$ [mag] & 17.61 & 15.16 & 16.44 & 16.32 & 0.59 \\ 
M$_{r}$ [mag] & -16.52 & -18.69 & -17.40 & -17.26 & 0.57 \\ 
$\mu_{0_{r}}$ [mag arcsec$^{-2}$] & 21.71 & 6.64 & 16.63 & 16.98 & 3.15 \\
$g-r$ [mag] & 1.61 & 0.27 & 0.59 & 0.57 & 0.24 \\
$\nabla (g-r)$ [mag dex$^{-1}$] & 0.38 & -0.47 & -0.05 & -0.07 & 0.20 \\
\textbf{Late Types} & & & & & \\
n$_{r}$ & 8.15 & 0.50 & 1.38 & 1.15 & 0.94 \\
$R_{e_{r}}$ [arcsec] & 12.49 & 0.15 & 2.91 & 2.58 & 1.79 \\
m$_{r}$ [mag] & 17.51 & 12.28 & 16.14 & 16.24 & 0.88 \\ 
M$_{r}$ [mag] & -10.63 & -19.27 & -17.42 & -17.53 & 1.02 \\ 
$\mu_{0_{r}}$ [mag arcsec$^{-2}$] & 22.20 & 6.30 & 19.89 & 20.23 & 2.00 \\
$g-r$ [mag] & 0.71 & 0.20 & 0.44 & 0.44 & 0.10 \\
$\nabla (g-r)$ [mag dex$^{-1}$] & 0.66 & -1.67 & -0.06 & -0.06 & 0.26 \\
\hline  
\hline  
    \end{tabular}
\caption{Statistical parameters of the photometric properties in the $r$ band of the MaNDala sample and for the $g-r$ colors and color gradients. The statistics for the early and late type sub samples are also presented for the 135 galaxies for which we have morphological information.
}
\label{table_phot}
\end{table}


Using the ellipticity profiles derived from our analysis, we can interpolate the $\epsilon$ value at any radius in the $r$ band. In particular we can also interpolate an approximation of the radius that contains $90\%$ of the light ($r_{90}$), using the cumulative flux profiles for each galaxy (upper right panel in Figure \ref{fig:manga-8145-3702-profiles}), and then interpolate the $\epsilon$ at that radius for all the galaxies in the sample. Assuming these values as the overall ellipticities for the galaxies, they can be converted into inclinations ($\cos i = b/a$). The MaNDala sample has galaxies with various inclinations, with a mean of $53.83^{\circ}$ ($\epsilon_{mean}$ = 0.43); 18 galaxies ($\sim 13\%$ of the sample) are highly inclined, $i > 70^{\circ}$ ($\epsilon\sim$ 0.66 ), while 14 ($\sim 10\%$ of the sample) exhibit low inclinations, $i < 30^{\circ}$ ($\epsilon\sim$ 0.13). Recall that cuts were applied to derive all the profiles, including the cumulative flux one, which means that the $r_{90}$ may be underestimated.

\begin{figure}
\centering
\subfloat{%
  \includegraphics[width=1\columnwidth]{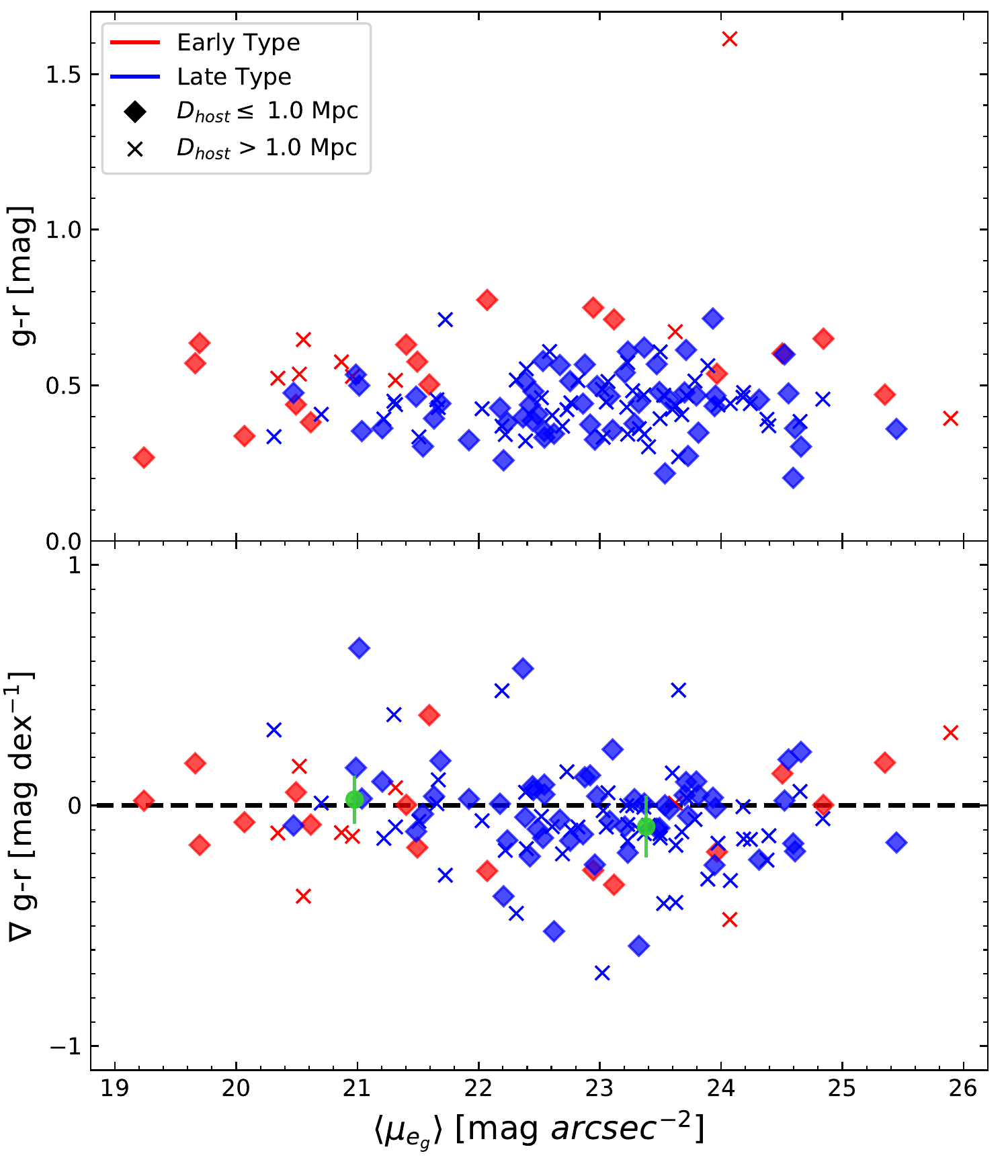}%
}
\caption{Color and color gradients in the top and bottom panels respectively. Both panels are color coded according to morphology types, red for early-types and blue for late-types. Symbols correspond to environment, where diamonds represent galaxies with $D_{host}\le$ 1.0 Mpc and crosses galaxies with $D_{host}>$ 1.0 Mpc. Green points in the bottom panels represent the mean $\langle\mu_{e}\rangle_{g}$ and $\nabla (g-r)$ values for all galaxies below and above 22 mag $arcsec^{-2}$ in $\langle\mu_{e}\rangle_{g}$, error bars represent their standard deviation.}
\label{fig:color_gradcolor_VSmueffmean}
\end{figure}

 In Figure \ref{fig:color_gradcolor_VSmueffmean} we show the relation between the mean effective SB \citep{GrahamDriver05}, which is defined as:

    \begin{equation} 
        \langle\mu_{e}\rangle = m + 2.5log_{10}(2\pi R_{e}^{2}),
                        \label{eq:mean_eff_SB}
    \end{equation}
and the $g-r$ color and the color gradient $\nabla (g-r)$ as well. To calculate $\langle\mu_{e}\rangle_{g}$ we use our estimations of the Sersic apparent magnitude in $g$, and the $R_{e}$ estimated in the $g$ band.
There is no significant trend of the $g-r$ color with $\langle\mu_{e}\rangle_{g}$, neither for ETGs nor for LTGs. We use crosses and diamonds to indicate field and group dwarfs, respectively. There is no clear segregation in this plot due to environmental characterization, however, the highest SB dwarfs in the sample are all ETGs in denser enviroments, $D_{\rm host}\le 1$ Mpc. Regarding the $g-r$ gradient, for the dwarfs with $\langle\mu_{e}\rangle_{g}<22$ (high SBs) are more common the positive gradients, while for those with lower SB's are more common the flat or negative gradients. There is no clear segregation due to environment.

\begin{figure}
\centering
\subfloat{%
  \includegraphics[width=1\columnwidth]{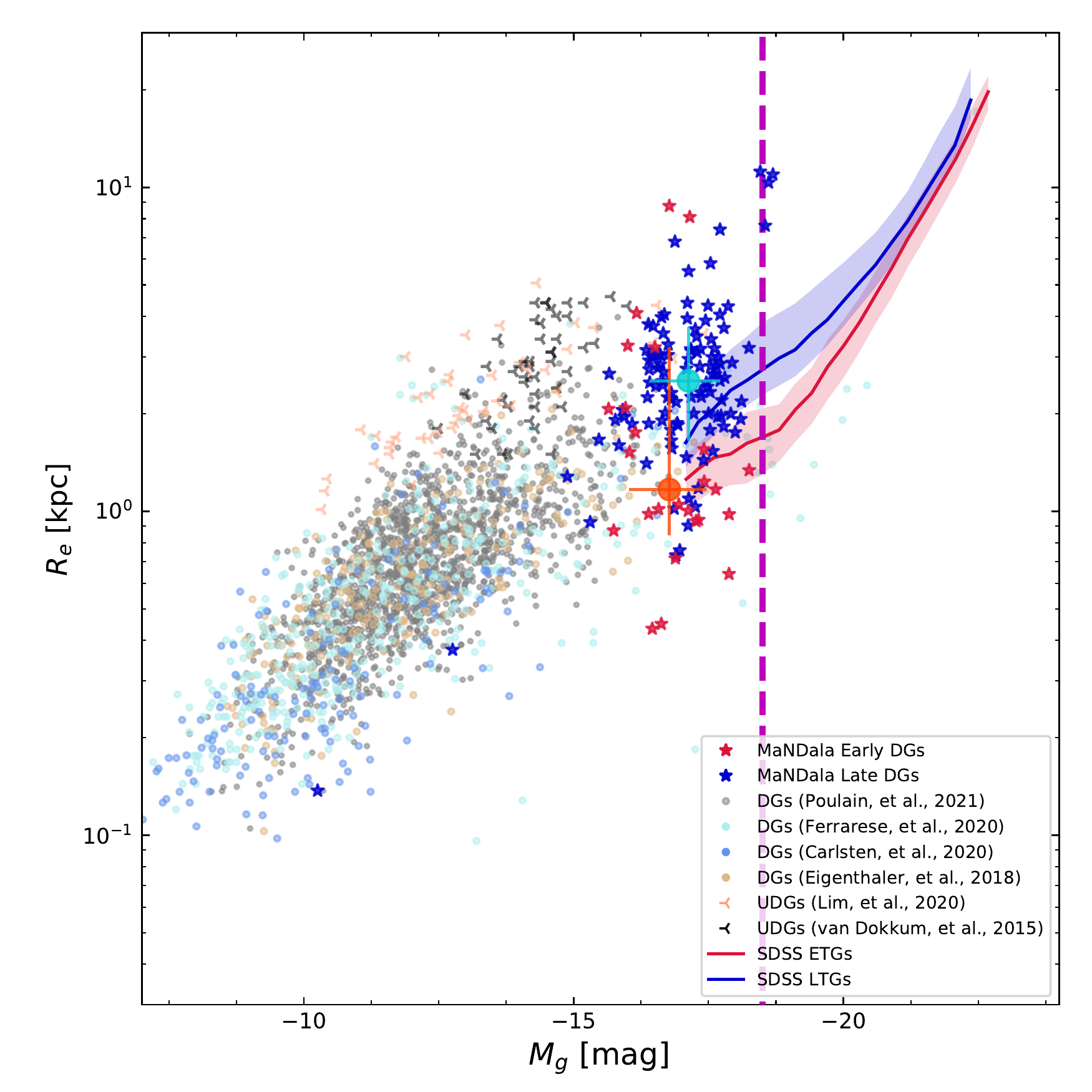}%
}
\caption{The $g$-band absolute magnitude vs. $g$-band effective radius of the MaNDala sample, separated into LTG and ETG dwarfs (blue and red stars, respectively). The large blue and red circles with error bars are the respective running median and 16-84th  percentiles. For comparison purposes, other samples of DGs are plotted (see the sources in the inset) The magenta dashed line marks the M$_{g}$ = -18.5 mag limit imposed to select this sample, used with the NSA Catalog photometric data.. 
\label{fig:MgVsRe_Comparison}}
\end{figure}

A classical way to characterize dwarfs is using the so-called Kormendy relations \citep{Kormendy1977,Kormendy1985}, which relate galaxy parameters that can be inferred by a S\'ersic fit. In Figure \ref{fig:MgVsRe_Comparison} we show one of these relations, which compares the absolute magnitude in the $g$ band and $R_{e}$ in the same band. In this plot we aim to locate the MaNDala sample within a summary similar to the one presented by \citet{Poulain2021}, which compares several DGs samples \citep{Poulain2021,Ferrarese2020,Carlsten2020,Eigenthaler2018}, Ultra Difuse Galaxies (UDGs) \citep{Lim2020,vanDokkum2015}.
Along with these samples we present results for normal (giant) early and late type galaxies represented with solid red and blue lines. Effective radii were obtained from the \citet{Meert+2013,Meert+2016} catalogs based on the SDSS DR7 and the morphologies from the  \citet{Huertas-Company+2011} catalog. We compute these relations for galaxies with spectroscopic redshifts $z\leq0.07$ in order to avoid biases due to the PSF resolution of the SDSS. The $g$ band magnitudes were K-corrected and Evolution-corrected to $z=0$ following \citet{Dragomir+2018} and \citet{Rodriguez-Puebla+2020a}. We see that our sample populates the brighter end of the distribution for DGs, indicating that these objects are indeed bright DGs, with the exception of a few galaxies that fall below and within the cloud of DGs for magnitudes fainter than $-16$. For a given $M_g$, the radii of our early-type dwarfs agree with the low-luminosity end of the \citet{Meert+2013,Meert+2016} catalogs of normal ETGs, though the former have a very large scatter. For our late-type dwarfs, they tend, on average, to have larger radii than the low-luminosity end of normal LTGs.
 In particular, there is a fraction of MaNDala galaxies with very large radii, $R_e>3$ kpc, for their luminosities. They appear as an extension to higher magnitudes of the UDCs depicted in this Figure. In general, our results show that the $R_e-M_g$ relation tends to flatten, specially for LTGs, in the range $-19\le M_g< -15$, though with a large scatter.


\begin{figure}
\centering
\subfloat{%
  \includegraphics[width=1\columnwidth]{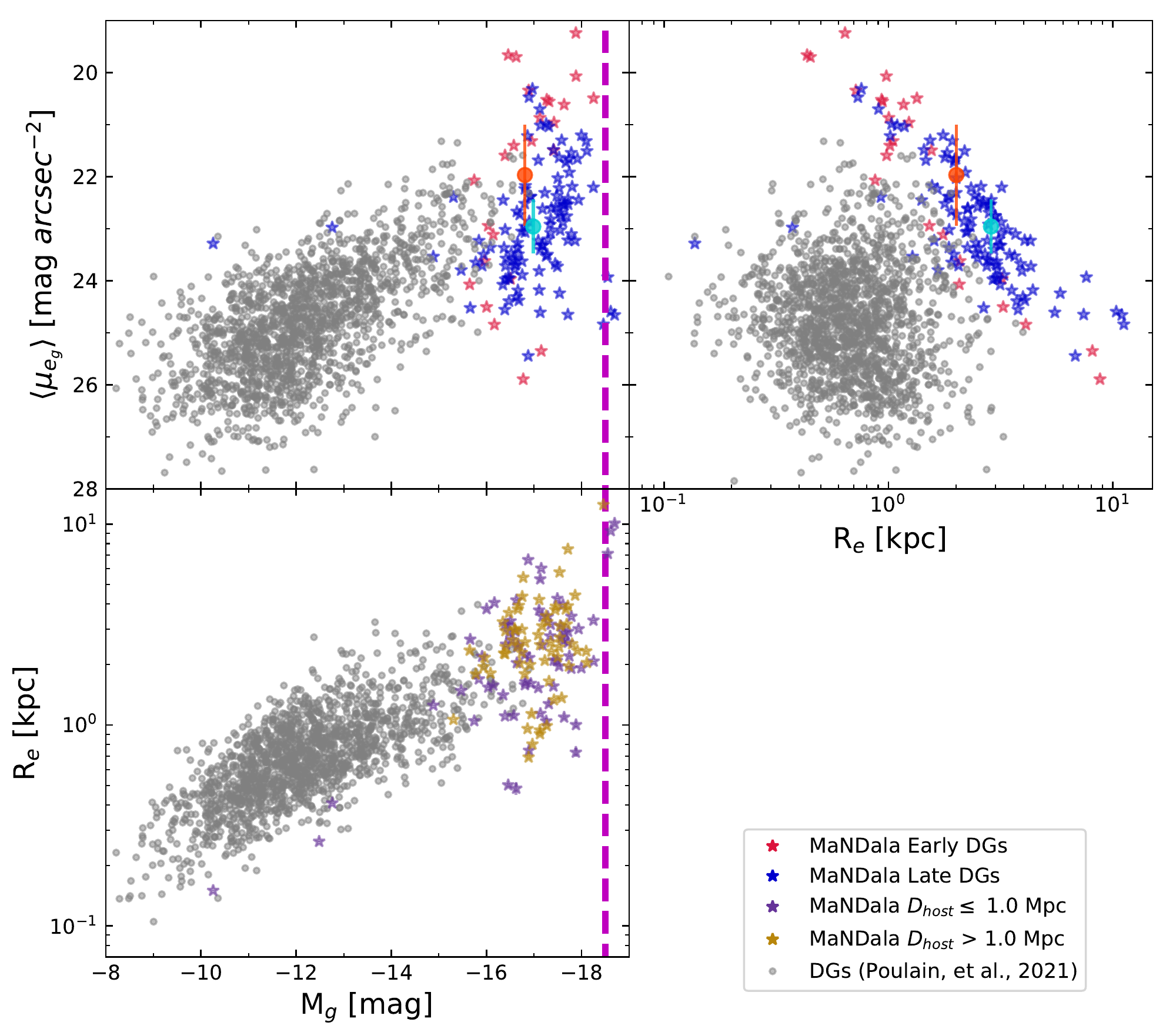}%
}
\caption{Kormendy's diagrams for the MaNDala sample in the $g$ band, compared with the \citet{Poulain2021} dwarf galaxy sample. In the upper panels our galaxies are color-coded to represent the two morphology groups present in the sample: red for early-types and blue for late-types, while the red and blue circles represent their means, the error bars are their respective $1\sigma$. In the bottom panel the color code represents the environment: dark yellow for dwarfs with $D_{host} \leq$ 1.0 Mpc and purple for those with $D_{host} >$ 1.0 Mpc. In the left side panels magenta dashed line marks the M$_{g}$ = -18.5 mag limit imposed to select this sample, used with the NSA Catalog photometric data.
\label{fig:Kormendy_diagram}}
\end{figure}

In Figure \ref{fig:Kormendy_diagram} we present the full Kormendy's diagrams for the $g$ band. For comparison purposes, we present along with the MaNDala sample the one described in \citet[][]{Habas+2020} and \citet{Poulain2021}, with gray dots. In the upper panels the relations between $\langle\mu_{e}\rangle$ and absolute magnitude $M_g$ (left panel) and $R_{e}$ (right panel) are presented. In the left plot, in general terms, the mean SB increases for brighter galaxies, which is visible as our sample continues the tendency drawn by the \citet{Poulain2021} sample. However, those low SB galaxies with $\langle\mu_{e,g}\rangle\gtrsim 24.5$ mag arcsec$^{-2}$ are outliers in this relation. These galaxies may be candidates for so-called UDGs\footnote{\citet{Poulain2021}, following the definition of UDGs by \citet[][]{vanDokkum2015}, find that the $g$-band effective SB of UDGs are larger than $\approx 24.5$ mag arcsec$^{-2}$.} and we have already noticed them in the $R_e-\langle\mu_{e}\rangle$ diagram shown in Figure \ref{fig:MgVsRe_Comparison}. Note also that our dwarf ETGs have, on average, smaller radii and higher mean effective SB's than the dwarf LTGs, while their magnitudes are similar. As for the right side plot, most of the MaNDala galaxies lie in a diagonal band, where for larger $R_e$, the $\langle\mu_{e,g}\rangle$ becomes lower. According to Eqn. (\ref{eq:mean_eff_SB}), this is due to the short range in absolute magnitudes of our sample.
However, comparing the MaNDala sample against a larger and less luminous sample, such as the one given by \citet{Poulain2021}, makes it easier to understand that MaNDala is an extension of this sample to larger/brighter dwarfs. In any case, there are dwarfs that seem to be outliers from the main trend, those with $\langle\mu_{e,g}\rangle\gtrsim 24.5$ mag arcsec$^{-2}$ and $R_e\gtrsim 1.5$ kpc (these are roughly the criteria to define UDGs, \citealp{Poulain2021,vanDokkum2015}). These galaxies also seem to be outliers in the $M_g-\langle\mu_{e}\rangle$ and $M_g-R_e$ diagrams, and as mentioned above, they may be UDG candidates. Note also that there are a few MaNDala galaxies that strongly deviate from the main trend, but at the other extreme, those with high SB's and small radii for their luminosity, i.e., very compact objects. These are mainly ETG dwarfs.


We can analyze the details of the individual galaxies that lay in the two already identified outlier groups. Starting with the UDG candidates, we find twelve galaxies that fulfill the afore mentioned conditions ($\langle\mu_{e,g}\rangle\gtrsim 24.5$ mag arcsec$^{-2}$ and $R_e\gtrsim 1.5$ kpc). However, five of them have reduced $\chi^{2}$ values for their $g$ band S\'ersic fit larger than 2.5; this leads to think that their location in the Kormendy diagram may be inaccurate. Another galaxy of this group exhibits a small $\chi^{2}$ value of its $g$ band fit (of 0.6), which may also indicate that this fit may not be optimal. The remaining six galaxies, could be considered as UDGs candidates. Four of them, manga-8487-9101, manga-9494-6103, manga-9876-12704, and 10517-12704, exhibit 
early-type morphologies, however the last two show a clearly bright center. The other two, manga-10221-12704 and manga-10841-12705, have late type morphologies. However, the last one shows signs of interaction. On the other hand, to select the MaNDala galaxies with large $\langle\mu_{e}\rangle$ values and small radii, we impose an arbitrary limit of $\langle\mu_{e}\rangle \leq$ 20 $mag$ $arcsec^{-2}$. Only three galaxies surpass this limit, however one of them also have $\chi^{2}$ values for their $g$ band S\'ersic fit larger than 2.5. For the remaining two (manga-8727-3702 and manga 9495-1901), their $\chi^{2}$ values are close to unity ($\sim$ 0.8) and we can assume that are indeed very compact and bright DGs.

The lower panel of Figure \ref{fig:Kormendy_diagram} is similar to Figure \ref{fig:MgVsRe_Comparison}, but indicating now the field/group information for our sample. As can be seen, there is not a clear segregation of the MaNDala galaxies by this environmental characterization in the $M_g-R_e$ diagram. We have verified that this segregation does not appear in the other diagrams either. We also do not observe a notable segregation between central and satellite dwarfs.

\begin{figure}
\centering
\subfloat{%
  \includegraphics[width=1\columnwidth]{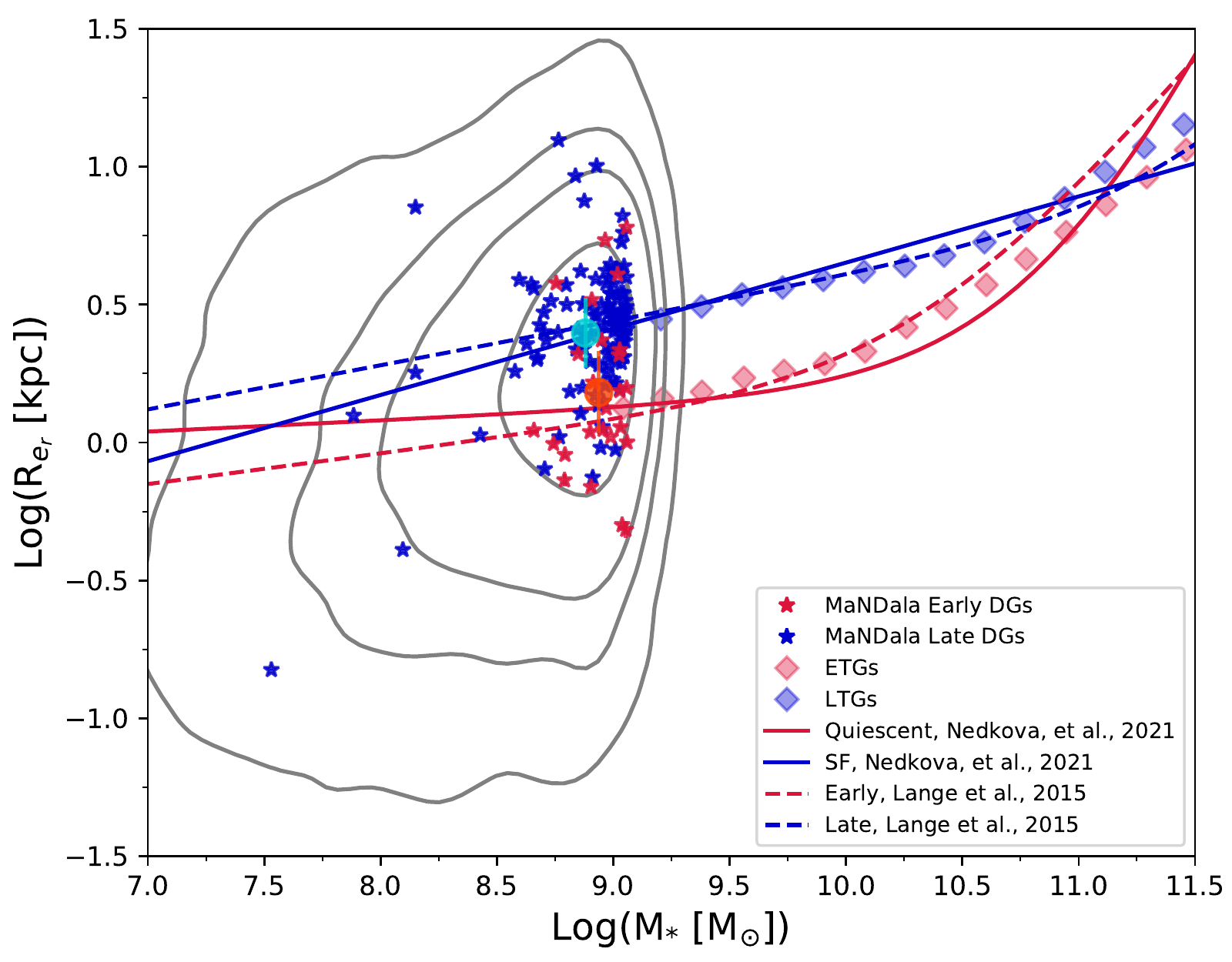}%
}
\caption{Location of the MaNDala galaxies in the mass-size relation (in red and blue stars for ETGs and LTGs respectively, their averages are shown by the red and blue circles, while the error bar shows $1\sigma$ of their distribution). For comparison, the contours underneath trace the location of all the NSA DGs, that fulfill the same selection criteria as our sample (from the innermost to the outermost they represent: 0.5$\sigma$, 1$\sigma$, 1.5$\sigma$ and 2$\sigma$ of this data set). The solid lines are the fits reported for this relation by \citet{Nedkova2021} for quiescent and star forming galaxies in red and blue respectively. The dashed lines are also fits, reported by \citet{Lange2015} for early and late type galaxies respectively.  Diamonds in red an blue are a comparison for higher mass galaxies with derivations for the $R_{e}$ given by \citet{Meert+2013,Meert+2016} and for the $M_{*}$ by \citet{Rodriguez-Puebla+2020a}.}
\label{fig:Mass_Size}
\end{figure}

Finally, in Figure \ref{fig:Mass_Size} we locate the MaNDala galaxies in the $M_{*}$ - $R_{e}$ diagram, shown as stars (color coded according to their morphological type as in Figures \ref{fig:MgVsRe_Comparison} and \ref{fig:Kormendy_diagram}), along with all the DGs in the NSA Catalog, selected with the same criteria as the ones in our sample. The resulting 25,998 NSA galaxies are traced by grey contours. 
The vast majority of the MaNDala galaxies are within the 1-$\sigma$ distribution of the NSA Catalog. In solid lines we show two of the mass-size relation fits reported by \citet{Nedkova2021}, corresponding to quiescent and star-forming galaxies in red and blue respectively (in Section \ref{Results_Spectroscopy} we will show that almost all the LTG dwarfs are star forming). Both fits were derived for galaxies in a redshift range of 0.2 $< z <$ 0.5, and for the entire range of $M_{*}$ considered in their sample ($10^{7}M_{\odot} < M_{*} < 10^{11.6}M_{\odot}$). We also show the fits provided by \citet{Lange2015} for ETGs and LTGs from the GAMA local survey, using their morphology cut, in red and blue dashed lines, respectively. In addition, we compare to the $M_{*}$ - $R_{e}$ relationship derived from the effective radii reported in \citet{Meert+2013,Meert+2016} catalogs (as in Figure \ref{fig:MgVsRe_Comparison}) and the stellar masses for this survey as derived in \citet{Rodriguez-Puebla+2020a}. All the plotted mean relations agree well with our results for the MaNDala galaxies, both for LTGs and ETGs. On average, the former are larger than the latter at a given stellar mass.

\subsection{Spectroscopic Results}
\label{Results_Spectroscopy}

In this Section we present results of global properties related to the stellar populations and emission lines of the MaNDala galaxies, obtained from the spectroscopic analysis described in \S\S \ref{Analysis_spectra}. The galaxy properties displayed in this Section are available as part of the SDSS-IV MaNDala VAC (see the details in Appendix \ref{AppendixVAC}). We leave for a series of future articles to explore the evolutionary and spatially resolved results of the fossil record analysis that can be derived using the MaNGA data. 


\begin{figure}
\centering
\includegraphics[width=1\columnwidth]{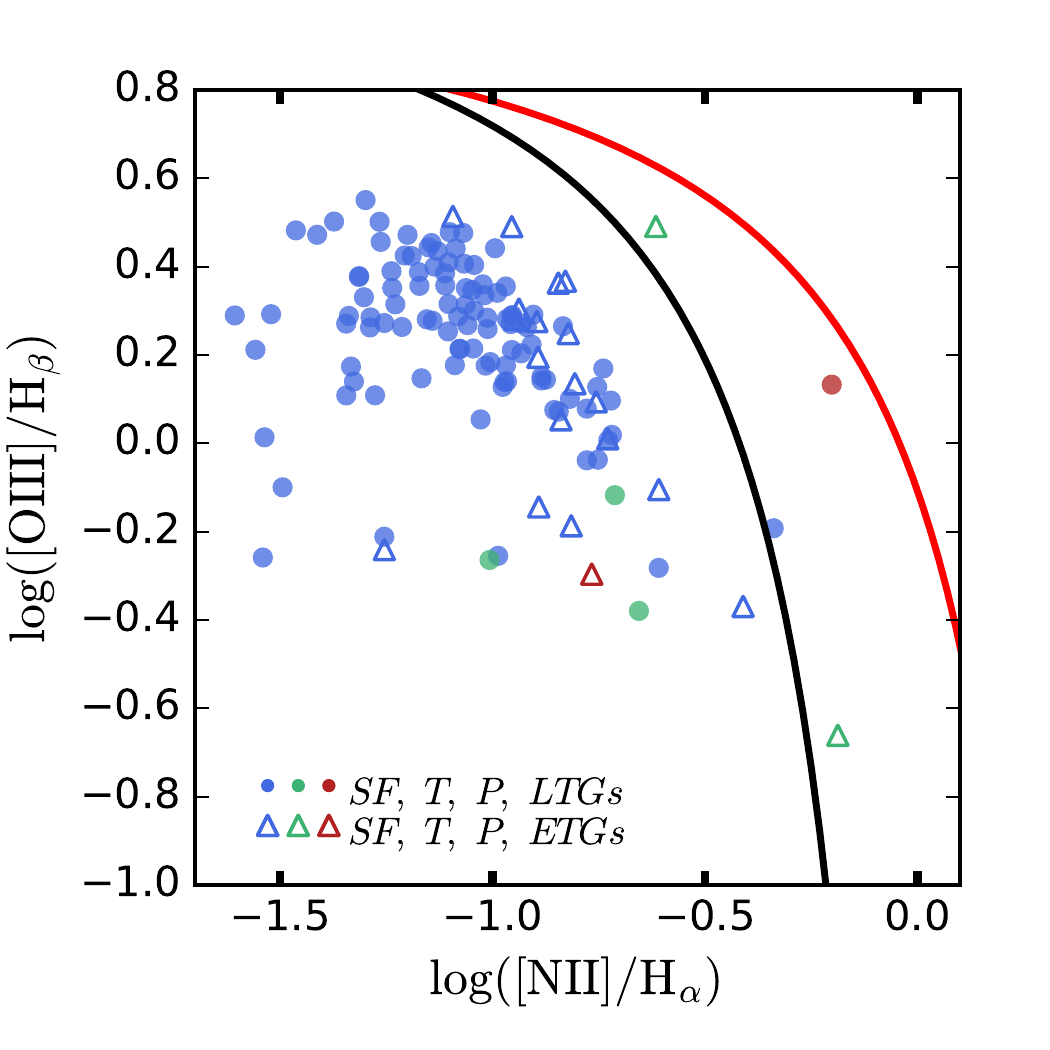}%
\caption{The BPT-NII diagram for the MaNDala galaxies reported in this paper. The line intensities are integrated within the FoV of each galaxy. The blue, green, and red colors refer to our classification into Star Forming (SFg), Transitioning (T), and Passive (P) galaxies, respectively, while closed circles and open triangles are for the LTG and ETG dwarfs, respectively. The lines indicate different criteria to identify the level of SF and AGN activity: red line indicates the Kewley criteria \citep{Kewley+2001}, black line indicates the Kauffmann criterion \citep{Kauffmann+2003May}.}
\label{fig:BTP}
\end{figure}

\subsubsection{Ionized gas results}

In Figure \ref{fig:BTP} we present the Baldwin, Philips \& Terlevich NII diagram \citep[BPT;][]{BPT1981} for the MaNDAla galaxies, using the corresponding emission line fluxes integrated within the FoV of each galaxy. The BPT diagram, as well as other line diagnostic diagrams \citep[][]{VeilleuxOsterbrock1987,Kewley+2001}, are used to distinguish the ionization mechanisms of nebular gas, which can be associated to active galacti nuclei (AGN), SF or hot old stars.
Based on the BPT-NII diagram and on the {\sc H}$\alpha$ EW, we attempt a classification of the SF activity level of MaNDala galaxies, following  the criteria discussed in \citet{Sanchez+2014}, \citet{Cano-Diaz+2016}, and  \citet{Cano-Diaz+2019}. Three types of galaxies are defined:
\begin{enumerate}
    \item  Star Forming (SFg), those with EW({\sc H}$\alpha$)$>6\AA$ and below the Kewley line \citep[black solid line in Fig. \ref{fig:BTP};][]{Kewley+2001};
    \item Passive (P; referred to as quiescent, quenched or retired as well), those with EW({\sc H}$\alpha$)$<3\AA$ independently of their position in the BPT diagram; and
    \item Transitioning (T), those with $3\AA\le$EW({\sc H}$\alpha$)$\le 6\AA$. 
\end{enumerate}

Applying the above criteria, we find that 115 galaxies ($92\%$) are SFg, 5 are P galaxies ($4\%$), and 5 are T ($4\%$). The blue, red, and green symbols in Figure \ref{fig:BTP} show the SFg, P and T galaxies respectively, while closed circles and open triangles are for the LTG and ETG dwarfs, respectively. In Figure \ref{fig:BTP}, only 2 P galaxies are plotted. This is because the H$\alpha$ and [OIII] lines of the other 3 P galaxies are very low, so that that the line ratios for them are well outside the ranges of the axes. Definitively, the great majority of our DGs in the MaNGA sample have signatures of being SFg. Among the dwarf LTGs, 96\% are SFg, 2\% are T and 2\% are P. For the dwarf ETGs, these fractions are 75\%, 12.5\%, and 12.5\%, accordingly. Thus, even for the ETGs, most of the dwarfs are SFg galaxies. On the other hand, we did not find significant signatures of AGN in any MaNDala galaxy when using the BPT diagram, however further and detailed analysis is required to exclude the possibility of finding nuclear activity in any MaNDala galaxy as previous studies of dwarf galaxies observed in MaNGA have find AGN signatures \citep[][with samples selected to have galaxies with: $M_{*}<$ 3 and 5 $\times 10^{9}$ $M_{\odot}$ respectively]{Mezcua2020,Penny2018}.
Summarizing, most of dwarf galaxies in the MaNGA survey are SFg without signatures of AGN contribution. In Section \ref{sect:sfhs_mandala}, we will explore the SFR-\ms\ relationship and show that our DG sample is a low-mass extension of the Main Sequence of SFg galaxies.

\subsubsection{Properties of the Stellar Populations}
\label{sec:SPS-results}

We use the (mass- and luminosity-weighted) stellar age and metallicity maps based on the SPS analysis from {\sc pyPipe3D} described in \S\S \ref{Analysis_spectra} to estimate the respective (mass- and luminosity-weighted) integrated ages and metallicities: \agemw, \agelw, \zmw, and \zlw, for each MaNDala galaxy. We also use the index maps to obtain the integrated {\sc D4000} index, defined as the ratio between the continuum flux within $4050-4250\AA$ and $3750-3950\AA$ \citep{Sanchez16b}. In the VAC, the above quantities are reported within the FOV and $R_e$ for each individual galaxy, but in the present work, unless otherwise is specified, we present these results within the FOV.

\begin{figure*}
\centering
\gridline{\fig{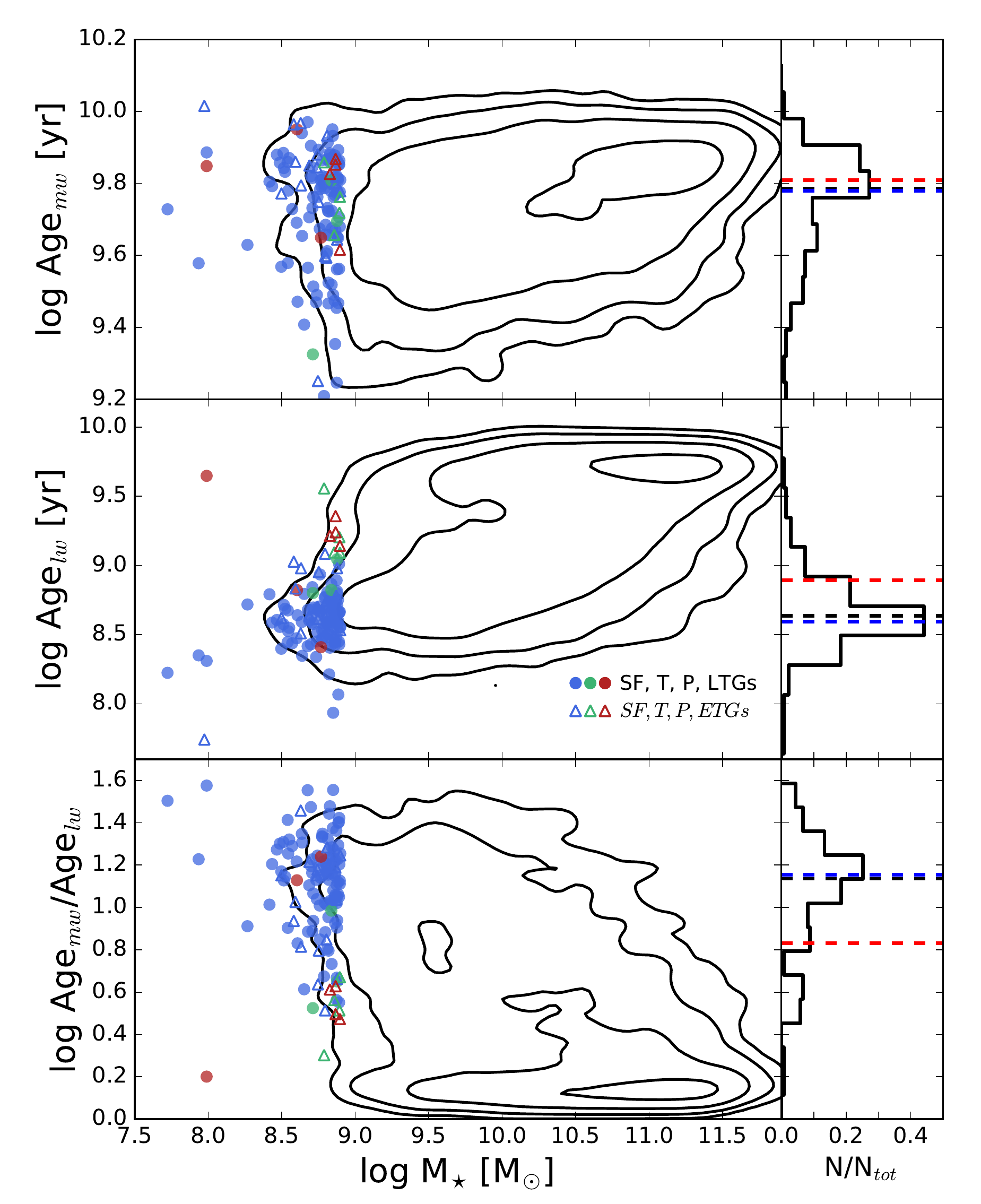}{0.5\textwidth}{}
          \fig{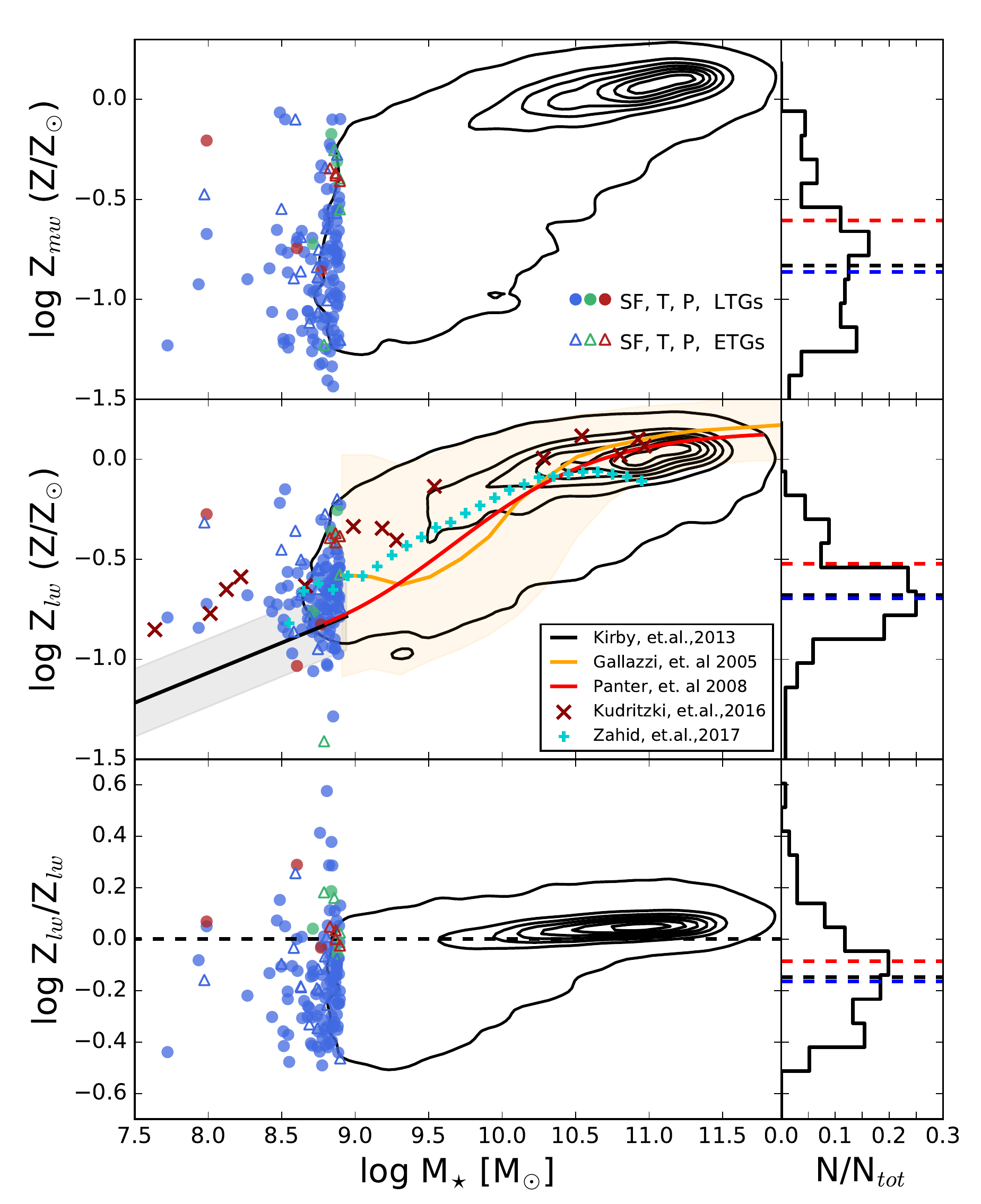}{0.5\textwidth}{}
        }
\caption{Left panels: Stellar mass--age relations for the MaNDala galaxies (symbols) and the whole MaNGa sample (iso-density contours). The ages are defined within the FoV of each galaxy, while for \ms, we used the NSA catalog. The upper and medium panels show the mass- and luminosity-weighted ages, respectively, while the lower panel shows the logarithm of their ratio. The colors and symbols are as in Figure \ref{fig:BTP}. The isocountors enclose $99\%$, $96\%$, $84\%$, and $37\%$ of the data (no correction for volume completeness was applied). Their right inclined panels show the distributions of $\log\agemw$, $\log\agelw$, and $\log(\agemw/\agelw)$ for the MaNDala galaxies. The black dashed lines indicate the respective medians, while the blue and red dashed lines correspond to medians of the LTG and ETG subsamples, respectively. Right panels: Same as left panels but for the mass- and light-weighted stellar metallicities. The isocontours in this case enclose $97\%$, $61\%$,  $45\%$, $33\%$, $23\%$, $13\%$, and $5\%$ of the data (no correction for volume completeness was applied). In the middle panel, we have added determinations from several previous studies, indicated in the inset box (see text). \label{fig:AgeAndMet_mass}}
\end{figure*}

The two upper left panels of Figure \ref{fig:AgeAndMet_mass} show \agemw\ and \agelw\ vs. \ms, respectively, for our DG sample (averaging of the SSP's ages is logarithmic, see \S\S \ref{Analysis_spectra}). The colors and symbols are as in Figure \ref{fig:BTP}. The isodensity contours in the same figure present the results for the whole MaNGA sample. As for \ms, we use the masses from the NSA catalog. The masses calculated with {\sc pyPipe3D} from the MaNGA data cubes are, on average, slightly lower than those from the NSA catalog, after passing to the Chabrier IMF; this may be due to aperture effects. For a more detailed discussion, see Sanchez et al. in prep.


According to the left panels in Figure \ref{fig:AgeAndMet_mass}, low-mass galaxies, and dwarfs in particular, show a wide range of mass-weighted ages, the latter with a median of $6.2$ Gyr and 16th-84th percentiles of 7.1--4.2 Gyr; see  the right panel for the full distribution. On average, these ages are slightly lower than those of the massive galaxies in the MaNGA survey. The situation is different for the light-weighted age: dwarfs have much lower ages than massive galaxies and with a small scatter. The median and 16th-84th percentiles of \agelw\ are 0.4 and 0.6--0.3 Gyrs, respectively. The few dwarfs with $\agelw>1$ Gyr are passive or in transition. Unlike \agemw, which has a similar distribution for dwarf LTGs and ETGs, \agelw\ for dwarf ETGs tend to have larger values than for dwarf LTGs. The lower left panel of Figure \ref{fig:AgeAndMet_mass} shows the logarithmic difference of mass- and light-weighted ages for the MaNDala and the whole MaNGA samples. While for massive galaxies, this difference is on average small, with a median of $0.18$ dex, for the dwarfs, the differences are very large, with a median of $1.15$ dex, and with a large scatter; see the right panel for the full distribution. The differences are smaller on average for dwarf ETGs than for the LTG ones. The mass-weighted age is biased towards older stellar population, which informs about when a significant fraction of the stellar mass was formed, while the light-weighted age is more sensitive to late episodes of SF. 
Therefore, the difference between \agelw\ and \agemw\ is indicative of how coeval or dispersed in time the SFH of a galaxy was or it can also indicate the presence of very recent bursts of SF, as is the case of post starburst galaxies \citep{Plauchu-Frayn+12,Lacerna+20,Ge+2021}. If the difference is very large, it points out to two markedly different populations in the SSP decomposition of the observed spectrum: one that formed early and is dominant in the total stellar mass, and another associated to late SF episodes, with a low contribution to the total mass. With age differences as large as factors of 3--30, this seems to be the case for the MaNGA dwarfs. 

In the right panels of Figure \ref{fig:AgeAndMet_mass} we show the stellar mass--metallicity relationships, both for \zmw\ and \zlw, for the dwarf and whole MaNGA samples; the right panels show the respective metallicity distributions. Both the mass- and light-weighted stellar metallicities of dwarfs are significantly lower than those of massive galaxies, as many previous works have shown \citep[e.g.,][]{Ikuta+02,Hidalgo+17,McQuinn+20}; the medians (16th and 84th percentiles) of log($\zlw/Z_\odot$) and log($\zmw/Z_\odot$) are $-0.82$ ($-0.54,-1.17$) and $-0.67$ ($-0.47, -0.84$), respectively. This points to a decreasing in the chemical enrichment with a decrease in the stellar mass of the galaxies. DGs also show a large scatter in stellar metallicities, specially in the mass-weighed one. There is a weak preference for higher metallicities in dwarf ETGs than dwarf LTGs.
It is interesting that, on average, \zlw{} is larger than \zmw{} for dwarfs (though with a large scatter, including some galaxies with even an inverse result), while for the massive galaxies \zlw{} is mostly slightly smaller than \zmw. The median (16th and 84th percentiles) of $\log$(\zmw/\zlw) for dwarfs is $-0.15$ ($0.03,-0.34$) dex. The above results suggest that the young stellar populations of most of the dwarfs formed from gas that underwent more chemical enrichment (probably this is re-accreted gas) than the pristine gas from which the old populations formed. In contrast, for massive galaxies the opposite is valid, that is, their younger stellar populations were formed from less metallic gas, probably because these systems have accreted important amounts of mostly pristine gas over their lifetimes and their old populations were significantly enriched by early SF.

In the middle right panel of Figure \ref{fig:AgeAndMet_mass} we plot several previous measures of \zlw\ for dwarf and normal galaxies. When necessary, we homogenize to our adopted value of $Z_\odot=0.019$. The solid black line and grey shaded area show the mean relation and its scatter obtained by \citet[][]{Kirby+2013} from observations of resolved stars in Local Group dwarfs, both spheroidals and irregulars. The red crosses also correspond to determinations from the spectra of resolved stars, in this case, in local normal galaxies and DGs \citep[][and more references therein]{Kudritzki+2016}. The works of \citet[][solid orange line]{Gallazzi+2005} and \citet[][solid red line]{Panter+2008} used spectral information from the SDSS and applied different analysis techniques and SPS methods for determining \zlw. In the case of \citet[][blue crosses]{Zahid+2017}, stacked spectra of only SFg were used. In general, the MaNGA \zlw--\ms\ relation obtained with pyPipe3D is in rough agreement with previous studies. As for the dwarfs, our \zlw\ determinations are on average higher than those of \citet[][]{Kirby+2013} and in good agreement with those of \citet[][]{Zahid+2017} and \citet[][]{Kudritzki+2016}.



\begin{figure}
\centering
\subfloat{%
  \includegraphics[width=1\columnwidth]{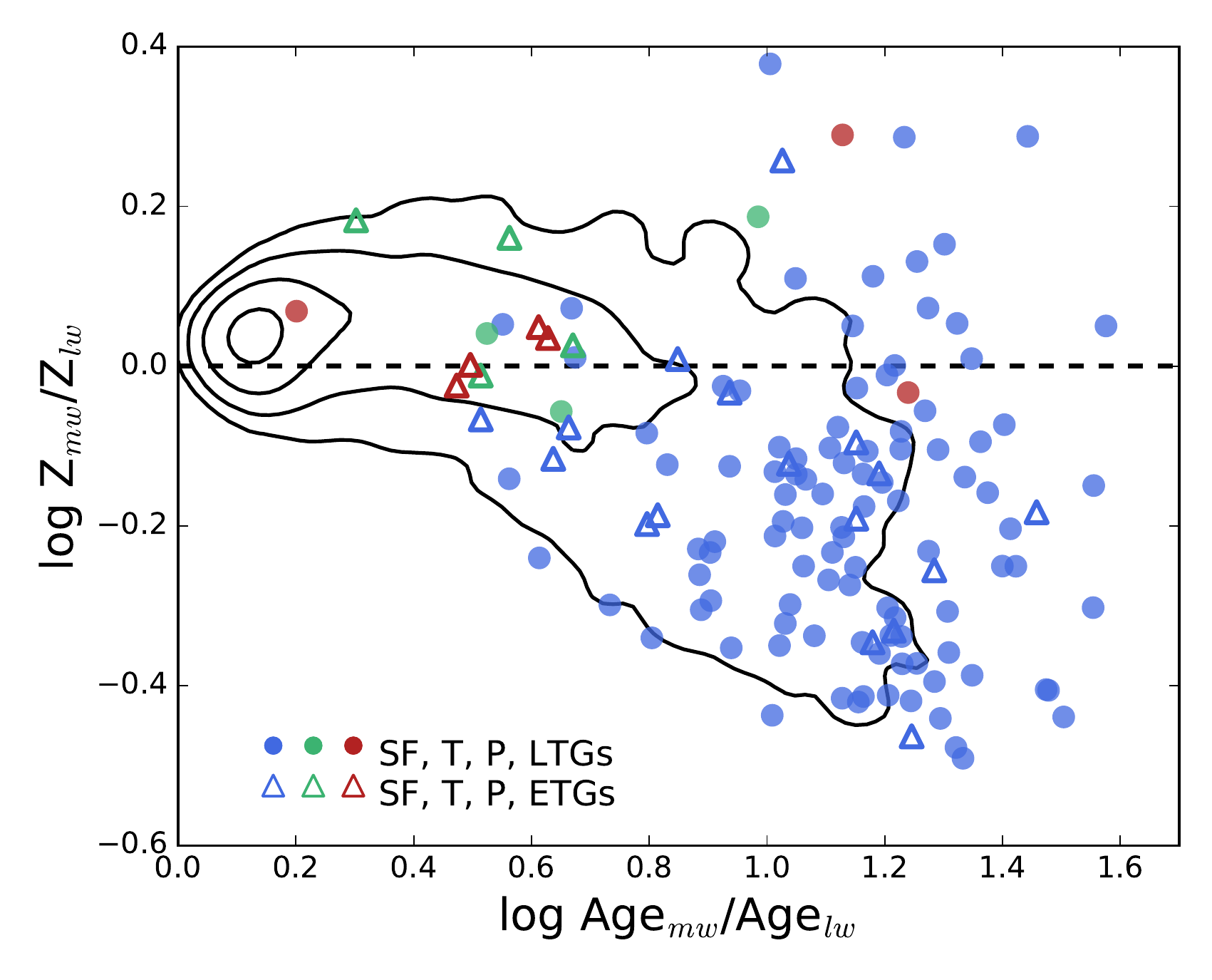}%
}
\caption{Relationship between the \agemw/\agelw\ and \zmw/\zlw\ ratios showed in the lower panels of Figure \ref{fig:AgeAndMet_mass} for the MaNDAla (symbols) and whole MaNGA (isodensity contours) samples. Symbols and colors are as in Figure \ref{fig:AgeAndMet_mass}. The isocountors enclose $99\%$, $96\%$, $84\%$, and $37\%$ of the data. }

\label{fig:Met_Age}
\end{figure}

In Figure \ref{fig:Met_Age}, we show the relationship between the \agemw/\agelw\ and  \zmw/\zlw\ ratios for the dwarf and whole MaNGA samples. The loci of DGs in this diagram clearly differ from the loci of the most massive galaxies. As discussed above, for dwarfs, on average, $\agemw>>\agelw$ and $\zmw<\zlw$, such that they tend to lie in the lower right side of this diagram, but with a large scatter. 
The trend seen in Figure \ref{fig:Met_Age} suggests a diversity of stellar populations for dwarfs, but in most of the cases they are characterized by a dominant old population with low metallicity and the presence of late, chemically enriched populations. The above points to a diversity of (bursty) SFHs, but characterized, on average, by an intense early phase of transformation of low-metallicity gas into stars and late episodes of SF from chemically enriched gas (the latter suggests a poor or absent contribution of pristine gas accretion during the late evolution of dwarfs). The larger the time interval between early and late burst of SF episodes, the more enriched is, on average, the gas from which the young populations form. 

For massive galaxies, the small values of the \agemw/\agelw\ and  \zmw/\zlw\ ratios, point to more continuous and homogeneous SFHs, and the fact that \zmw/\zlw\ tends to be slightly larger than 1, suggests that most of the gas out of which form younger stellar populations is of cosmological origin, that is, it is accreted from the poorly enriched intergalactic medium as the dark matter halo grows. As it is well known, within the $\Lambda$CDM cosmogony, the more massive the haloes, the later is their mass growth \citep[see e.g.,][]{Mo+2010,Rodriguez-Puebla+2016}, including the accretion of baryons.
Following this trend, the growth of low-mass haloes, those that host DGs, is very slow at late epochs. Therefore, most of the gas out of which young populations form in dwarfs galaxies is expected to come from inside the same halo or galaxy; it is (enriched) gas that was likely  heated/ejected by early SF in the galaxy that lately gets cold and falls back again. The results presented above agree with this general picture.

Finally, we have explored whether the age-- and metallicity--mass relations of our DGs segregate by being central or satellites or being at distances $D_{\rm host}$ larger or smaller than 1 Mpc. We do not find any clear trend in both cases.


\subsubsection{Properties of the global SFH}
\label{sect:sfhs_mandala}

\begin{figure}
\centering
\subfloat{%
  \includegraphics[width=1\columnwidth]{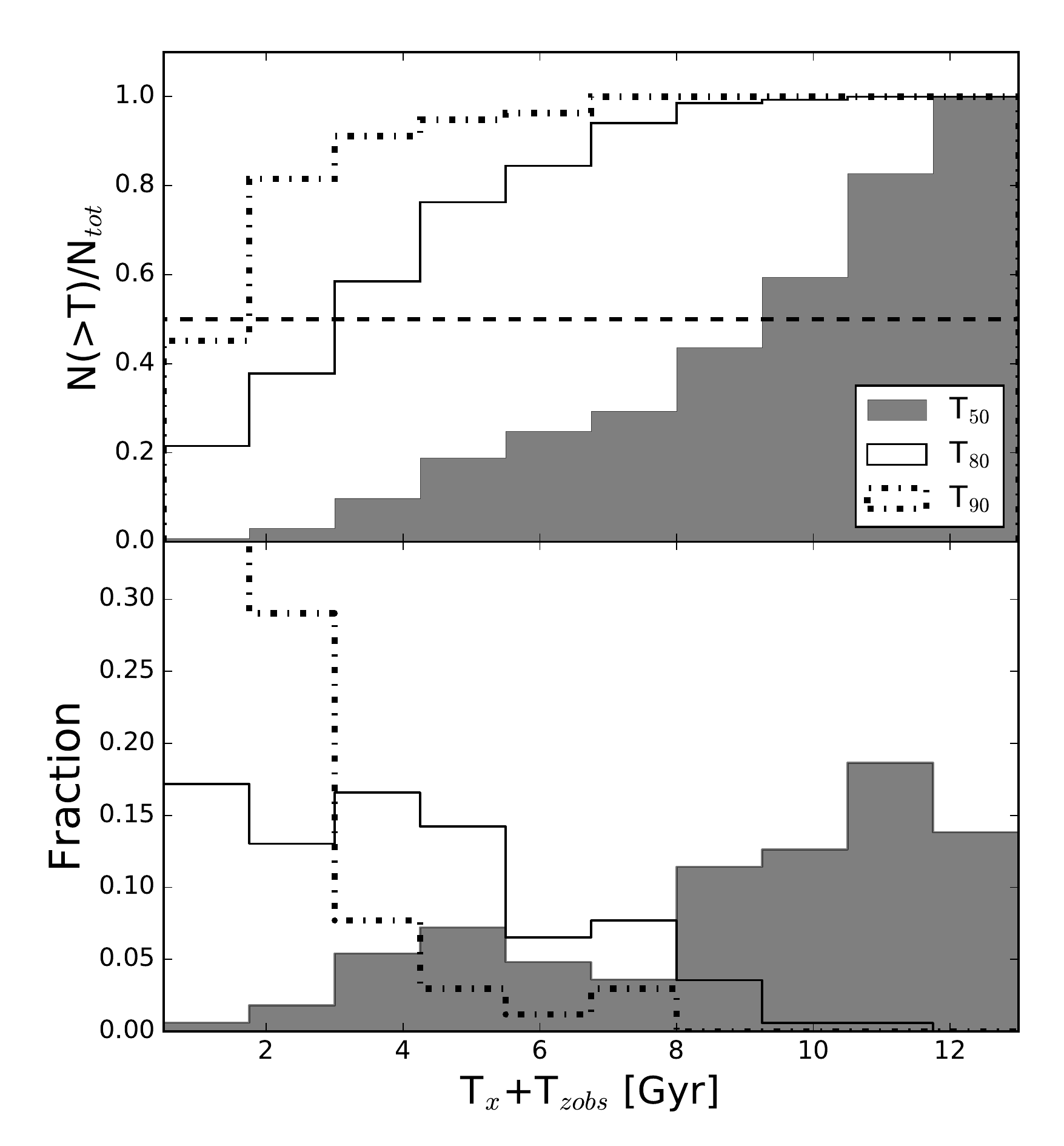}%
}
\caption{Cumulative (upper panel) and differential (lower panel) distributions of the look-back times at which 50\%, 80\%, and 90\% of the stellar mass of the MaNDala galaxies were formed. $T_{x}$, with $x=$ 50\%, 80\%, and 90\%, is the age at which the given fraction of stars were formed and $T_{z_{\rm obs}}$ is the look-back time at the observation redshift of a given galaxy for the assumed cosmology in this paper. The dashed line of the upper panel marks the 0.5 value for the cumulative distribution.}
\label{fig:T50}
\end{figure}

The SPS analysis of the MaNGA galaxies allows us to reconstruct their full global/local SF, stellar mass, and chemical enrichment histories \citep[see e.g.,][]{Ibarra-Medel+16,Goddard+2017,Rowlands+2018,Sanchez+2019,Peterken+2021}. In a forthcoming paper, we will present and discuss results related to these histories for the DGs. By using the global stellar mass growth history of each galaxy, we report in the VAC the stellar ages when 50\% ($T_{50}$) and 90\% ($T_{90}$) of the final stellar mass was formed respectively. Since most of the targets in the MaNDAla catalog are at very low redshifts $z<0.03$ (look-back time less than 0.4 Gyr for the cosmology adopted here), the stellar ages mentioned above differ from the respective cosmic look-back times only by small amounts of time. In any case, we calculate the look-back time at which each galaxy formed a given fraction of its \ms\ by adding to the respective stellar age (e.g., $T_{50}$) the look-back time of observation, $T_{z_{\rm obs}}$. 
Figure \ref{fig:T50} presents the differential and cumulative histograms of the look-back times (for the  cosmology adopted here) at which the MaNGA DGs formed 50\%, 80\%, and 90\% of their stellar masses. From the distributions, we see that most of the dwarfs formed half of their stars at very early epochs. For 50\% (70\%) of them, these epochs correspond to look-back times larger than $\approx 10$ (8) Gyr, that is, $z\gtrsim 1.9$ ($z\gtrsim 1.1$).
On the other hand, the formation of the last 20\% or 10\% of stars in the MaNGA dwarfs happened at late cosmic times. The medians of the look-back times corresponding to 80\% and 90\% of the formed stars are $3.7$ and $1.8$ Gyr, respectively, that is, half of the dwarfs, formed their last 20\%  (10\%) stars at $z\lesssim 0.33$ ($z\lesssim 0.14$).
The above results are in rough agreement with those by \citet{Zhou+2020}, who also analized a sample of MaNGA low-mass galaxies by using a Bayessian spectrum parametric fitting.

\begin{figure}
\centering
\subfloat{%
  \includegraphics[width=1\columnwidth]{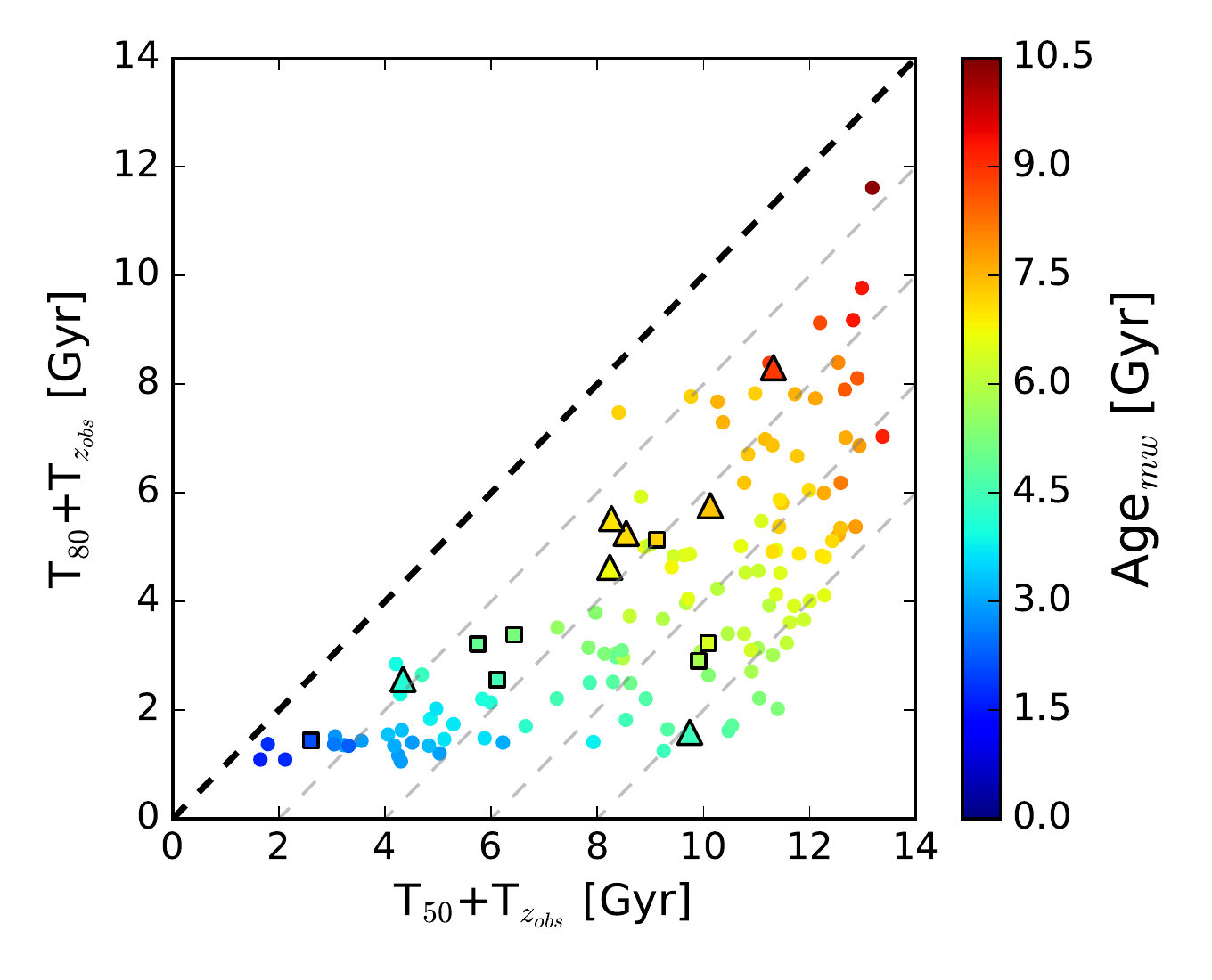}%
}
\caption{Relationship between the look-back times at which 50\% and 80\% of the stellar mass of the MaNDala galaxies were formed. The color of the symbols represents the mass-weighted ages, \agemw. The half-mass look-back time correlates well with \agemw.  Circles, squares and triangles correspond to SFg, T and P galaxies. The diagonal thin dashed lines indicate differences between the 50\% and 80\% look-back times of 2, 4, 6 and 8 Gyrs.}
\label{fig:T50-T80}
\end{figure}

An interesting question, related to the shape of the SFHs, is how different are the look-back times at which two different stellar mass fractions were formed. In Figure \ref{fig:T50-T80} we show the look-back times corresponding to 50\% and 80\% of the formed stars for the dwarfs. The color indicates the mass-weighted age of each galaxy. First, as expected, there is a good correlation between \agemw\ and the half-mass formation time, though the former tends to be shorter than the latter by $1-3$ Gyr, on average for ages large than $\sim 4$ Gyr. According to Figure \ref{fig:T50-T80}, the dwarfs that formed 50\% of their stars late ($\approx 2-5$ Gyr ago or $\agemw\approx 1-4$ Gyr), formed the last 20\% of stars very late, $1-1.5$ Gyr ago. For the oldest galaxies, there is a weak trend of earlier formation of the last 20\% of stars as earlier is the formation of half of the stars. The galaxies with higher differences between the formation of the 50\% and 80\% of their stellar masses ($>6$ Gyr) are those with intermedium ages, $\agemw\approx 5-8$ Gyr. The dwarfs classified as passive, P (triangles) present small differences between the formation of the 50\% and 80\% of their stellar masses, 2-4 Gyr. Since these galaxies are retired, they are not expected to have late SF events, able to contribute to a late growth of the last 20\% of stellar mass.

\begin{figure}
\centering
\subfloat{%
  \includegraphics[width=1\columnwidth]{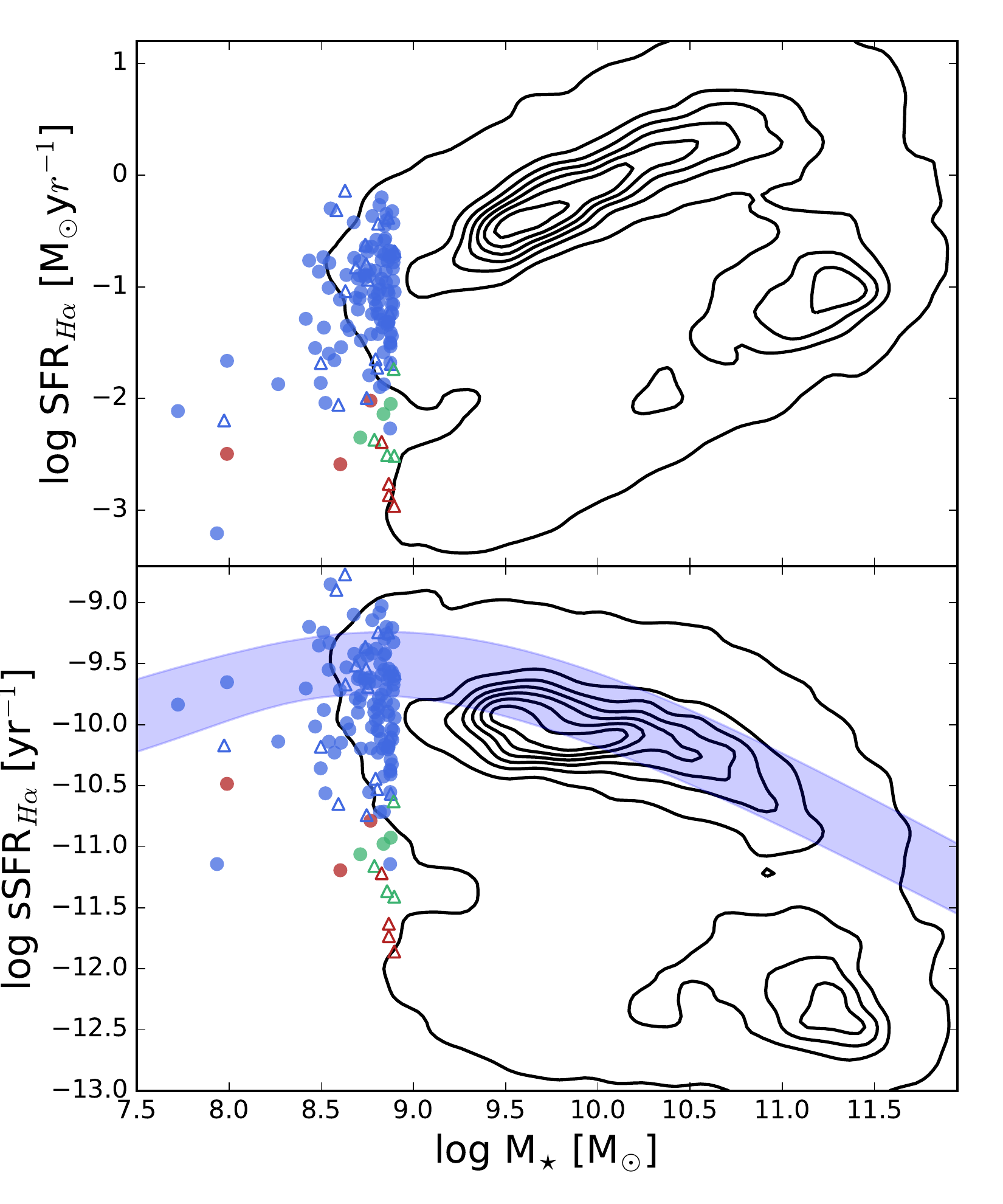}%
} 
\caption{{\it Upper panel:} Global  $SFR$ vs. \ms\ for the MaNDala (circles) and whole MaNGA (isodensity contours) samples. The $SFR$'s were calculated from the {\sc H}$\alpha$ line integrated within the FoV of each galaxy using the Chabrier IMF (see text). The stellar masses are from the NSA catalog. No correction for volume completeness was applied. {\it Lower panel:} Same as in the upper panel but for the specific SFR. The shaded area is the 1-$\sigma$ region describing the SF main sequence inferred from several observational data at $z\sim0.1$ by \citet{Rodriguez-Puebla+2020}.} 
\label{fig:SFMS}
\end{figure}

Finally, in the upper panel of Figure \ref{fig:SFMS}, we present the global $SFR$ vs. \ms\ relationship for the MaNDala and the whole MaNGA samples. We have calculated the SFR from the  dust-corrected {\sc H}$\alpha$ line integrated within the FoV of each galaxy, using the Kennicutt conversion factor corrected to the \citet{Chabrier} IMF. In addition, we introduce here a nebular metallicity-dependent correction to this factor. For this, we follow \citet[][]{Shin+2021} and use the pyPipe3D nebular metallicity calculated for each MaNDala galaxy. We also apply a correction to our {\sc H}$\alpha$-based SFR, which for dwarfs is systematically underpredicted with respect to the FUV-based SFR \citep[][]{Lee+2009}, likely because at low SFRs
statistically sampling the IMF does not produce enough massive OB
stars, resulting in a deficit for {\sc H}$\alpha$ measurements \citep[][]{Weidner+2013}. This correction is applied following \citet[][]{Shin+2021} and it is actually very small for most of our dwarfs. 

Dwarf SFg's (blue circles) follow roughly the trend towards low masses of the main sequence of galaxies of the whole sample, though with a large scatter. The few dwarfs classified as P and T (red and green colors, respectively), as expected, lie much below the SF main sequence.  
The lower panel of Figure \ref{fig:SFMS} shows the sSFR ($=SFR/\ms$) vs. \ms.    
The data at lower masses is scarce but they hint to {\it a bending in the specific $SFR-\ms$ relation at masses around $10^9$ \msun}, in accordance with some previous observations and empirical inferences \citep[e.g.,][]{Skibba+2011,McGaugh+2017,Rodriguez-Puebla+2020}.  The local SF main sequence inferred by the latter authors from a large set of observations from the FUV to the FIR (obeying volume completeness) is over-plotted in the lower panel of Figure \ref{fig:SFMS}. In spite of the large differences in the methodologies and taking into account that the isocontours plotted in Figure  \ref{fig:SFMS} are not corrected by volume completeness, the agreement is reasonable, including its extension to dwarfs. 

\section{Summary and Discussion}\label{SummaryDiscussion}

In this work we present the sample of DGs  observed by the project MaNGA, for which we perform photometric and spectroscopic analyses, using DESI and MaNGA data respectively. The sample is conformed by 136 galaxies that were selected to have $M_{*} < 10^{9.06} M_{\odot}$ and $M_{g} > -18.5$\footnote{According to the stellar masses and absolute magnitudes from the NSA Catalog}, and it is to our knowledge the first large DGs sample observed with IFS data. 

The photometric analysis, carried out in the $g$, $r$ and $z$ bands provide SB profiles which were fitted with a S\'ersic function convolved with a Moffat function to describe the PSF of the photometric measurements. Due to the depth of the DESI images, the fitted SB profiles extend, on average, to $\sim 6, 4.5,$ and 3.5 $R_e$ in the $g, r$ and $z$ bands, respectively. Along with these, other radial profiles such as P.A., $\epsilon$ and color are obtained. The main results about this analysis are summarized as follows:

\begin{itemize}
  \item The galaxies conforming our sample, called MaNDala, are mainly bright dwarfs; this is particularly evident when compared with other DG samples, such as the one presented by \citet{Poulain2021}. The mean stellar mass of the sample is $10^{8.89} M_{\odot}$, while the mean absolute magnitude in $g$ is -16.92 mag. The sample is dominated by late-type and central galaxies.
  
  \item The S\'ersic index and the central and effective SBs of the dwarf ETGs tend to be higher and with a greater dispersion than for the dwarf LTGs. In general, the former are more compact and redder than the latter. Most MaNDala galaxies have nearly flat color profiles. 
  
  \item  The location of our galaxies within the $R_e-M_g$, $\langle\mu_{e,g}\rangle-M_g$, and $\langle\mu_{e,g}\rangle-R_e$ (Kormendy) diagrams, shows that this sample has a very large scatter in the implied relations. This makes visible the variety of MaNDala galaxies, which can range from very low SB and extended (UDG candidates) to high SB and compact.
  
  \item Late- and early-type DGs occupy roughly the same regions in the Kormendy diagrams, with a slight preference of the former to be more extended and of lower SB than the latter.

  \item MaNDala galaxies show a large scatter in the magnitude- or mass-size relations, consistent with NSA DGs or other DG samples, and confirm flattening in the  $-19\le M_g< -15$ or $10^8\le(\ms/\msun)<10^9$  ranges of these relations with respect to the more luminous/massive galaxies.
  
\end{itemize}

The spectroscopic analysis, performed in apertures containing the entire MaNGA FoV, made use of a non-parametric SPS analysis in spatial bins of $S/N > 50$ (pyPipe3D code). A single component Gaussian fit to the emission lines was afterwards performed. The main results from this analysis are summarized as follows:

\begin{itemize}

\item Using emission lines criteria, the dwarfs were classified into Star Forming (92\%), Transitioning (4\%), and Passive (4\%) galaxies; no conclusive signatures of nuclear activity were found in any of them. The vast majority of dwarf LTGs, 96\%, are SFg; even dwarf ETGs are mostly SFg,  75\%. 

\item The 16th-84th percentiles of \agemw\ and \agelw\ are 7.1-4.2 Gyr and 0.6-0.3 Gyr, respectively. The \agemw\ values of dwarfs are, on average, slightly smaller than those of massive galaxies, while these differences are much larger in the case of \agelw. As for the stellar metallicities of our dwarfs, they are much lower than those of massive galaxies, and present a large scatter; the 16th-84th percentiles of log($\zmw/Z_\odot$) and log($\zlw/Z_\odot$) are (-0.54,-1.17) and (-0.47,-0.84), respectively.

\item The loci of dwarfs in the \agemw/\agelw\ vs. \zmw/\zlw\ plane are different from massive galaxies. For dwarfs, on average, \agemw$>>$\agelw\ and \zmw$<$\zlw\ (with a large scatter), with a trend of a lower \zmw/\zlw\ ratio as \agemw/\agelw\ increases. 
The above points to a diversity of SFHs, but characterized on average, by an intense early phase of transformation of low-metallicity gas into stars and late episodes of SF from chemically enriched gas. The greater the time interval between the early burst and the final episodes of SF, the more enriched, on average, the gas from which young populations are formed is. 

\item Half of the dwarfs formed 50\% of their stellar mass at early epochs,
$z\gtrsim 2$. However, the formation of the last 20\% of the mass, happened recently ($\lesssim 0.3$ for half of the sample). The T and P dwarfs are those with less differences in their epochs of 50\% and 80\% stellar mass formation ($\sim 2-4$ Gyr), while SFg dwarfs with intermediate ages show differences larger than 6 Gyr, that is, their SFHs tend to have an early period of intense SF and very late SF episodes. 

\item The H$_\alpha$-based SFRs of our dwarfs present a large scatter. In the sSFR--\ms\ diagram, the SFg dwarfs seem to follow the SF main sequence of the more massive galaxies, but showing evidence of a bending of this relation at $\ms\sim 10^9$ \msun, with a median maximum value of log(sSFR/\msun yr$^{-1}$)$\approx -9.7$.

\end{itemize}

It should be said that we did not find significant differences between the central and satellite dwarfs or between the isolated and grouped dwarfs in any of the photometric/structural or stellar population relations presented here. The above may imply that the internal processes in the presence of a low gravitational potential are more relevant in shaping the properties of dwarfs than the external processes associated to the environment \citep[e.g.,][but see e.g., \citealp{Weisz+2011,Young+2014}]{Dunn+2015}.

Using new public data from deep photometric instruments, such as the data set provided by the DESI collaboration, gives the advantage to infer more precisely, not only the global characteristics of galaxies but also the true shape of the SB and geometric profiles of galaxies up to fainter SB values and to large radii, than the SDSS photometric instrument is able to provide. This may have an impact when investigating in detail the structure of these objects, but also when looking for restricting kinematical analyses, that require the use of precise geometrical parameters. 

On the other hand, the use of IFS observations, such as the ones MaNGA provides, allow us to access spatially resolved spectroscopic observations, which gives homogeneous information about their stellar and ionized gaseous components. In the near future we aim to extend our analysis to fully exploit the resolved nature of this data set, extending the integrated results presented here. 

This is the Accepted Manuscript version of an article accepted for publication in The Astronomical Journal. IOP Publishing Ltd is not responsible for any errors or omissions in this version of the manuscript or any version derived from it. This Accepted Manuscript is published under a CC BY licence. 

\section*{Acknowledgments}

Authors wish to acknowledge the comments given by the anonymous referee, which were very helpful to improve this paper. Authors also wish to acknowledge Claudia Castro Rodr\'iguez, for her help generating a fraction of the mosaic images for the MaNDala SDSS-IV VAC, and for designing and developing the project webpage.

MCD acknowledges support from CONACYT "Ciencia de Frontera" grant 320199. MCD and HMHT acknowledge support from UC MEXUS-CONACYT grant CN-17-128.
ARP acknowledge support from the CONACyT  `Ciencia Basica' 
grant 285721. EAO acknowledges support from the SECTEI (Secretaría de Educación, Ciencia, Tecnología e Innovación de la Ciudad de México) under the Postdoctoral Fellowship SECTEI/170/2021 and CM-SECTEI/303/2021.

Funding for the Sloan Digital Sky 
Survey IV has been provided by the 
Alfred P. Sloan Foundation, the U.S. 
Department of Energy Office of 
Science, and the Participating 
Institutions. 

SDSS-IV acknowledges support and 
resources from the Center for High 
Performance Computing  at the 
University of Utah. The SDSS 
website is www.sdss.org.

SDSS-IV is managed by the 
Astrophysical Research Consortium 
for the Participating Institutions 
of the SDSS Collaboration including 
the Brazilian Participation Group, 
the Carnegie Institution for Science, 
Carnegie Mellon University, Center for 
Astrophysics | Harvard \& 
Smithsonian, the Chilean Participation 
Group, the French Participation Group, 
Instituto de Astrof\'isica de 
Canarias, The Johns Hopkins 
University, Kavli Institute for the 
Physics and Mathematics of the 
Universe (IPMU) / University of 
Tokyo, the Korean Participation Group, 
Lawrence Berkeley National Laboratory, 
Leibniz Institut f\"ur Astrophysik 
Potsdam (AIP),  Max-Planck-Institut 
f\"ur Astronomie (MPIA Heidelberg), 
Max-Planck-Institut f\"ur 
Astrophysik (MPA Garching), 
Max-Planck-Institut f\"ur 
Extraterrestrische Physik (MPE), 
National Astronomical Observatories of 
China, New Mexico State University, 
New York University, University of 
Notre Dame, Observat\'ario 
Nacional / MCTI, The Ohio State 
University, Pennsylvania State 
University, Shanghai 
Astronomical Observatory, United 
Kingdom Participation Group, 
Universidad Nacional Aut\'onoma 
de M\'exico, University of Arizona, 
University of Colorado Boulder, 
University of Oxford, University of 
Portsmouth, University of Utah, 
University of Virginia, University 
of Washington, University of 
Wisconsin, Vanderbilt University, 
and Yale University.

This project makes use of the MaNGA-Pipe3D dataproducts. The dataproducts used in this project got benefit of computational and human resources provided by the LAMOD- UNAM project through the clusters Atocatl and Tochtli. LAMOD is a collaborative effort between the IA, ICN, and IQ institutes at UNAM and DGAPA UNAM grants PAPIIT IG101620 and IG10122  We thank the IA-UNAM MaNGA team for creating this catalog, and the CONACyT-180125 project for supporting them. 


%




\appendix

\section{The SDSS-IV MaNDala Value Added Catalog} \label{AppendixVAC}

All results presented in this work are available for public use as a Value Added Catalog (VAC) for the SDSS-IV Consortium named MaNDala. These results make up the first version of the MaNDala VAC (V1.0), and correspond to 136 galaxies of our sample. This version is being released as part of the SDSS DR17. A second version of this catalog is expected to be released in the future, and will be available from the same URLs given in this paper.

The data can be retrieved as a single Flexible Image Transport System\footnote{\url{https://fits.gsfc.nasa.gov/fits_documentation.html}} (FITS) format table named {\fontfamily{qcr}\selectfont mandala$\_$v1$\_$0.fits}, through the SDSS public repository\footnote{\url{https://data.sdss.org/sas/dr17/manga/mandala}}. In addition a collection of mosaic images in PDF format as the one presented in Figure \ref{fig:manga-8145-3702-mosaic}, are also available through the same site.

The documentation of the MaNDala VAC is available through the SDSS Value Added Catalogs web-page \footnote{\url{ https://www.sdss.org/dr17/data_access/value-added-catalogs/}}. Details of the MaNDala VAC FITS format file, coined as data model, are given in Table \ref{table_mandala_vac}. The same data model can also be found in the SDSS web page for this catalog\footnote{\url{https://data.sdss.org/datamodel/files/MANGA_MANDALA}}, and in our own site for this project\footnote{\url{https://mandalasample.wordpress.com}}, in which the reader may find extra information about the MaNDala project.

The file {\fontfamily{qcr}\selectfont mandala$\_$v1$\_$0.fits} consist of two extensions, a header (HDU0) containing information about the dataproducts from DESI and MaNGA used to derive our results, and the principal extension (HDU1), in which the results are stored in the form of a 84 columns table. Columns 1 to 54 in the MaNDala VAC FITS table contain the photometric results derived with the DESI data (see Section \ref{Data_desi}), which are described in Sections \ref{Analysis_Isophotes}, \ref{Analysis_Profiles}, \ref{Results_Sersic} and \ref{Results_Photometry}. Columns 55 to 84 contain spectroscopic results derived with MaNGA data (see Section \ref{Data_manga}), which are described in Sections \ref{Analysis_spectra} and \ref{Results_Spectroscopy}.

\clearpage
\LTcapwidth=\textwidth
\begin{longtable*}{ c c c c c }
\hline
\hline
Column No. & Name & Type & Units & Description \\
\hline
1 & Plateifu & String & & MaNGA Plate-IFU \\ 
2 & MangaID & String & & MaNGA ID \\ 
3 & R.A. & Float & deg & NSA Right Ascension (J200) \\ 
4 & Dec & Float & deg & NSA Declination (J2000) \\ 
5 & IAUName & String & & IAU Name \\ 
6 & NSA$\_$ID & Integer & & NSA ID \\ 
7 & NSA$\_$redshift & Float & & NSA redshift \\ 
8 & NSA$\_$LogSersicMass & Float & h$^{-2}$ M$\odot$ & Logarithm of NSA Sersic mass \\ 
9 & Radius & Float Array & arcsec & DESI Profiles radii \\ 
10 & SB$\_$g & Float Array & mag arcsec$^{-2}$ & g-band DESI Surface Brightness profiles \\ 
11 & SB$\_$g$\_$err & Float Array & mag arcsec$^{-2}$ & g-band DESI Surface Brightness profiles errors\\ 
12 & SB$\_$r & Float Array & mag arcsec$^{-2}$ & r-band DESI Surface Brightness profiles \\ 
13 & SB$\_$r$\_$err & Float Array & mag arcsec$^{-2}$ & r-band DESI Surface Brightness profiles errors \\ 
14 & SB$\_$z & Float Array & mag arcsec$^{-2}$ & z-band DESI Surface Brightness profiles \\ 
15 & SB$\_$z$\_$err & Float Array & mag arcsec$^{-2}$ & z-band DESI Surface Brightness profiles \\ 
16 & P.A$\_$r & Float Array & degrees & r-band DESI Position Angle profiles \\ 
17 & P.A$\_$r$\_$err & Float Array & degrees & r-band DESI Position Angle profiles errors \\
18 & Ellipticity$\_$r & Float Array & & r-band DESI Ellipticity profiles \\ 
19 & Ellipticity$\_$r$\_$err & Float Array & & r-band DESI Ellipticity profiles errors \\ 
20 & Flux$\_$r & Float Array & nanomaggies & r-band DESI Accumulated flux profile \\ 
21 & Interpolated$\_$Reff$\_r$ & Float & arcsec & r-band $R_{e}$ derived from the accumulated flux \\ 
22 & Interpolated$\_$R90$\_r$ & Float & arcsec & \shortstack{r-band Radius at 90\% of light \\ derived from the accumulated flux} \\ 
23 & Ellip$\_$R90$\_$r & Float & & r-band Ellipticity at Radius at 90\% of light \\ 
24 & P.A$\_$R90$\_$r & Float & degrees & r-band P.A. at Radius at 90\% of light \\ 
25 & Sersic$\_$SB$\_$eff$\_$g & Float & mag arcsec$^{-2}$ & g-band Sersic Surface Brightness at $R_{e}$ \\ 
26 & Sersic$\_$SB$\_$eff$\_$g$\_$err & Float & mag arcsec$^{-2}$ & g-band Sersic Surface Brightness at $R_{e}$ error\\ 
27 & Sersic$\_$Reff$\_$g & Float & arcsec & g-band Sersic $R_{e}$ \\ 
28 & Sersic$\_$Reff$\_$g$\_$err & Float & arcsec & g-band Sersic $R_{e}$ error \\ 
29 & n$\_$Sersic$\_$g & Float & & g-band n Sersic index \\ 
30 & n$\_$Sersic$\_$g$\_$err & Float & & g-band n Sersic index error \\ 
31 & Sersic$\_$SB$\_$0$\_$g & Float & mag arcsec$^{-2}$ & g-band Sersic central Surface Brightness \\ 
32 & Sersic$\_$AppMag$\_$g & Float & mag & g-band Sersic Apparent magnitude \\ 
33 & Sersic$\_$AbsMag$\_$g & Float & mag & g-band Sersic Absolute magnitude (h=1) \\ 
34 & Sersic$\_$Chi2$\_$g & Float & & g-band Reduced Chi$^{2}$ for Sersic fit \\ 
35 & Sersic$\_$SB$\_$eff$\_$r & Float & mag arcsec$^{-2}$ & r-band Sersic Surface Brightness at $R_{e}$ \\ 
36 & Sersic$\_$SB$\_$eff$\_$r$\_$err & Float & mag arcsec$^{-2}$ & r-band Sersic Surface Brightness at $R_{e}$ error\\ 
37 & Sersic$\_$Reff$\_$r & Float & arcsec & r-band Sersic $R_{e}$ \\ 
38 & Sersic$\_$Reff$\_$r$\_$err & Float & arcsec & r-band Sersic $R_{e}$ error \\ 
39 & n$\_$Sersic$\_$r & Float & & r-band n Sersic index \\ 
40 & n$\_$Sersic$\_$r$\_$err & Float & & r-band n Sersic index error \\ 
41 & Sersic$\_$SB$\_$0$\_$r & Float & mag arcsec$^{-2}$ & r-band Sersic central Surface Brightness \\ 
42 & Sersic$\_$AppMag$\_$r & Float & mag & r-band Sersic Apparent magnitude \\ 
43 & Sersic$\_$AbsMag$\_$r & Float & mag & r-band Sersic Absolute magnitude (h=1) \\ 
44 & Sersic$\_$Chi2$\_$r & Float & & r-band Reduced Chi$^{2}$ for Sersic fit \\ 
45 & Sersic$\_$SB$\_$eff$\_$z & Float & mag arcsec$^{-2}$ & z-band Sersic Surface Brightness at $R_{e}$ \\ 
46 & Sersic$\_$SB$\_$eff$\_$z$\_$err & Float & mag arcsec$^{-2}$ & z-band Sersic Surface Brightness at $R_{e}$ error\\ 
47 & Sersic$\_$Reff$\_$z & Float & arcsec & z-band Sersic $R_{e}$ \\ 
48 & Sersic$\_$Reff$\_$z$\_$err & Float & arcsec & z-band Sersic $R_{e}$ error \\ 
49 & n$\_$Sersic$\_$z & Float & & z-band n Sersic index \\ 
50 & n$\_$Sersic$\_$z$\_$err & Float & & z-band n Sersic index error \\ 
51 & Sersic$\_$SB$\_$0$\_$z & Float & mag arcsec$^{-2}$ & z-band Sersic central Surface Brightness \\ 
52 & Sersic$\_$AppMag$\_$z & Float & mag & z-band Sersic Apparent magnitude \\ 
53 & Sersic$\_$AbsMag$\_$z & Float & mag & z-band Sersic Absolute magnitude (h=1) \\ 
54 & Sersic$\_$Chi2$\_$z & Float & & z-band Reduced Chi$^{2}$ for Sersic fit \\ 
55 & Stellar$\_$mass$\_$FoV & Float & M$\odot$ & Log Stellar mass within the FoV$^{*}$ \\ 
56 & Stellar$\_$mass$\_$Reff & Float & M$\odot$ & Log Stellar mass within one $R_{e}^{**}$ \\ 
57 & SFRssp$\_$FoV & Float & M$\odot$ yr$^{-1}$ & Log SSP Star Formation Rate within the FoV$^{*}$ \\ 
58 & SFRssp$\_$Reff & Float & M$\odot$ yr$^{-1}$ & Log SSP Star Formation Rate within one $R_{e}^{**}$ \\ 
59 & T50$\_$FoV & Float & yr & \shortstack{Log formation time when the galaxy reached 50\% \\ of its total stellar mass (calculated within the FoV$^{*}$)} \\ 
60 & T50$\_$Re & Float & yr & \shortstack{Log formation time when the galaxy reached 50\% \\ of its total stellar mass (calculated within one $R_{e}^{**}$)} \\ 
61 & T90$\_$FoV & Float & yr & \shortstack{Log formation time when the galaxy reached 90\% \\ of its total stellar mass (calculated within the FoV$^{*}$)} \\ 
62 & T90$\_$Re & Float & yr & \shortstack{Log formation time when the galaxy reached 90\% \\ of its total stellar mass (calculated within one $R_{e}^{**}$)} \\ 
63 & D4000$\_$FoV & Float & dex & Average D4000 value defined within the FoV$^{*}$ \\ 
64 & D4000$\_$Reff & Float & dex & Average D4000 value defined within one $R_{e}^{**}$ \\ 
65 & Age$\_$lum$\_$FoV & Float & yr & Average Log Luminosity weighted age within the FoV$^{*}$ \\ 
66 & Age$\_$lum$\_$Reff & Float & yr & Average Log Luminosity weighted age within one $R_{e}^{**}$ \\ 
67 & Age$\_$mass$\_$FoV & Float & yr & Average Log Mass weighted age within the FoV$^{*}$ \\ 
68 & Age$\_$mass$\_$Reff & Float & yr & Average Log Mass weighted age within the $R_{e}^{**}$ \\ 
69 & Metallicity$\_$lum$\_$FoV & Float & ZH & \shortstack{Average Log Luminosity weighted metallicity \\ within the FoV$^{*}$} \\ 
70 & Metallicity$\_$lum$\_$Reff & Float & ZH & \shortstack{Average Log Luminosity weighted metallicity \\ within one $R_{e}^{**}$} \\ 
71 & Metallicity$\_$mass$\_$FoV & Float & ZH & \shortstack{Average Log Mass weighted metallicity \\ within the FoV$^{*}$} \\ 
72 & Metallicity$\_$mass$\_$Reff & Float & ZH & \shortstack{Average Log Mass weighted metallicity \\ within one $R_{e}^{**}$} \\
73 & SFRHa$\_$FoV & Float & M$\odot$ yr$^{-1}$ & Log H$\alpha$ Star Formation Rate within the FoV$^{*}$ \\ 
74 & SFRHa$\_$Reff & Float & M$\odot$ yr$^{-1}$ & Log H$\alpha$ Star Formation Rate within one $R_{e}^{**}$ \\ 
75 & Ha$\_$FoV & Float & erg s$^{-1}$ cm$^{-2}$ & Log H$\alpha$ flux within the FoV$^{*}$ \\ 
76 & Ha$\_$Reff & Float & erg s$^{-1}$ cm$^{-2}$ & Log H$\alpha$ flux within one $R_{e}^{*}$ \\ 
77 & Hb$\_$FoV & Float & erg s$^{-1}$ cm$^{-2}$ & Log H$\beta$ flux within the FoV$^{*}$ \\ 
78 & Hb$\_$Reff & Float & erg s$^{-1}$ cm$^{-2}$ & Log H$\beta$ flux within one $R_{e}^{**}$ \\ 
79 & NII$\_$FoV & Float & erg s$^{-1}$ cm$^{-2}$ & Log [NII]$_{6583}$ flux within the FoV$^{*}$ \\ 
80 & NII$\_$Reff & Float & erg s$^{-1}$ cm$^{-2}$ & Log [NII]$_{6583}$ flux within one $R_{e}^{**}$ \\ 
81 & OIII$\_$FoV & Float & erg s$^{-1}$ cm$^{-2}$ & Log [OIII]$_{5007}$ flux within the FoV$^{*}$ \\ 
82 & OIII$\_$Reff & Float & erg s$^{-1}$ cm$^{-2}$ & Log [OIII]$_{5007}$ flux within one $R_{e}^{**}$ \\ 
83 & EWHa$\_$FoV & Float & & Equivalent Width of H$\alpha$ within the FoV$^{*}$ \\ 
84 & EWHa$\_$Reff & Float & & Equivalent Width of H$\alpha$ within one $R_{e}^{**}$ \\ 
\hline  
\hline  
\caption{Data model of the MaNDala VAC FITS format table. $^{*}$Quantities given within the Field of View (FoV) are calculated within the MaNGA FoV. $^{**}$Quantities given within one effective radius (Reff) are calculated using our estimation of the $R_{e}$ from the Sersic fit (column 37).} 
\label{table_mandala_vac}
\end{longtable*}

\bibliography{biblio}
\bibliographystyle{aasjournal}



\end{document}